%% file: me.tex
\documentclass[times,twocolumn,final]{elsarticle}
\usepackage{framed,multirow}
\usepackage{amsmath}
\usepackage{amssymb}
\usepackage{latexsym}
\usepackage{url}
\usepackage{xcolor}
\usepackage{hyperref}
\usepackage{booktabs}
\usepackage{color}
\usepackage{graphicx}

\usepackage{bm}

\graphicspath{{./Figs/}}

\input{defs}

\definecolor{newcolor}{rgb}{.8,.349,.1}
\journal{arXiv}

\begin{document}
% \verso{M.~Wu \textit{et~al.}}
\begin{frontmatter} 
\title{Disentangled Latent Energy-Based Style Translation: An Image-Level Structural MRI Harmonization Framework} 

\author[1,2]{Mengqi {Wu}}
\author[1]{Lintao {Zhang}}
\author[1]{Pew-Thian {Yap}}
\author[3]{Hongtu {Zhu}}
\author[1]{Mingxia {Liu}\corref{cor1}}

\address[1]{Department of Radiology and Biomedical Research Imaging Center, University of North Carolina at Chapel Hill, 
Chapel Hill, NC 27599, USA
}
\address[2]{Joint Department of Biomedical Engineering, University of North Carolina at Chapel Hill and North Carolina State University, Chapel Hill, NC 27599, USA}
\address[3]{Department of Biostatistics and Biomedical Research Imaging Center, University of North Carolina at Chapel Hill, Chapel Hill, NC 27599, USA}
\cortext[cor1]{Corresponding author:  M.~Liu (mingxia\_liu@med.unc.edu).}

\begin{abstract}
Brain magnetic resonance imaging (MRI) has been extensively employed across clinical and research fields, but often exhibits sensitivity to \emph{site effects} arising from non-biological variations 
%non-biological variations arising from %site-specific factors site effects, 
such as differences in field strength and scanner vendors. % and imaging protocols. 
Numerous retrospective MRI harmonization techniques have demonstrated encouraging outcomes in reducing the site effects at image level. %{\color{red}at both the feature and image levels}.  
However, existing %image-level harmonization 
methods generally suffer from high computational requirements and limited generalizability, restricting their applicability to unseen MRIs. % data. 
%To address this, we introduce an image-level framework for structural MRI harmonization via disentangled latent energy-based style translation (DLEST). 
In this paper, we design a novel disentangled latent energy-based style translation (DLEST) framework for unpaired image-level MRI harmonization, consisting of (a) \emph{site-invariant image generation} (SIG), (b) \emph{site-specific style translation} (SST), and (c) \emph{site-specific MRI synthesis} (SMS). 
Specifically, %the DLEST is composed of three modules, 
the SIG employs a latent autoencoder to encode MRIs into a low-dimensional latent space and reconstruct MRIs from latent codes. %, irrespective of site labels. 
%This results in highly generalizable image generation. 
The SST utilizes an energy-based model to comprehend global latent distribution of a target domain and translate source latent codes toward the target domain, while SMS enables MRI synthesis with a target-specific style. 
By disentangling image generation and style translation in latent space, the DLEST can achieve %generalizable and 
efficient style translation. 
\if false
The SST module employs an energy-based model to comprehend global latent distribution of a target domain and translate source latent codes toward the target domain. This enables efficient latent-space style translation
% from multiple source domains to target 
and the third module enables MRI synthesis with a target-specific style. 
\fi 
Our model was trained on T1-weighted MRIs from a public dataset (with 3,984 subjects across 58 acquisition sites/settings) and validated on an independent dataset (with 9 traveling subjects scanned in 11 sites/settings) in four tasks: histogram and feature visualization, site classification, brain tissue segmentation, and site-specific structural MRI synthesis.
Qualitative and quantitative 
%Experimental 
results demonstrate the superiority of our method over several state-of-the-arts. % methods.  

\end{abstract}
\end{frontmatter}

%%%%%%%%%%Pipeline%%%%%%%%%
\begin{figure*}[t]
\setlength{\abovecaptionskip}{0pt}
\setlength{\belowcaptionskip}{-2pt}
\setlength{\abovedisplayskip}{-2pt}
\setlength{\belowdisplayskip}{-2pt}
\centering
\includegraphics[width=0.98\textwidth]{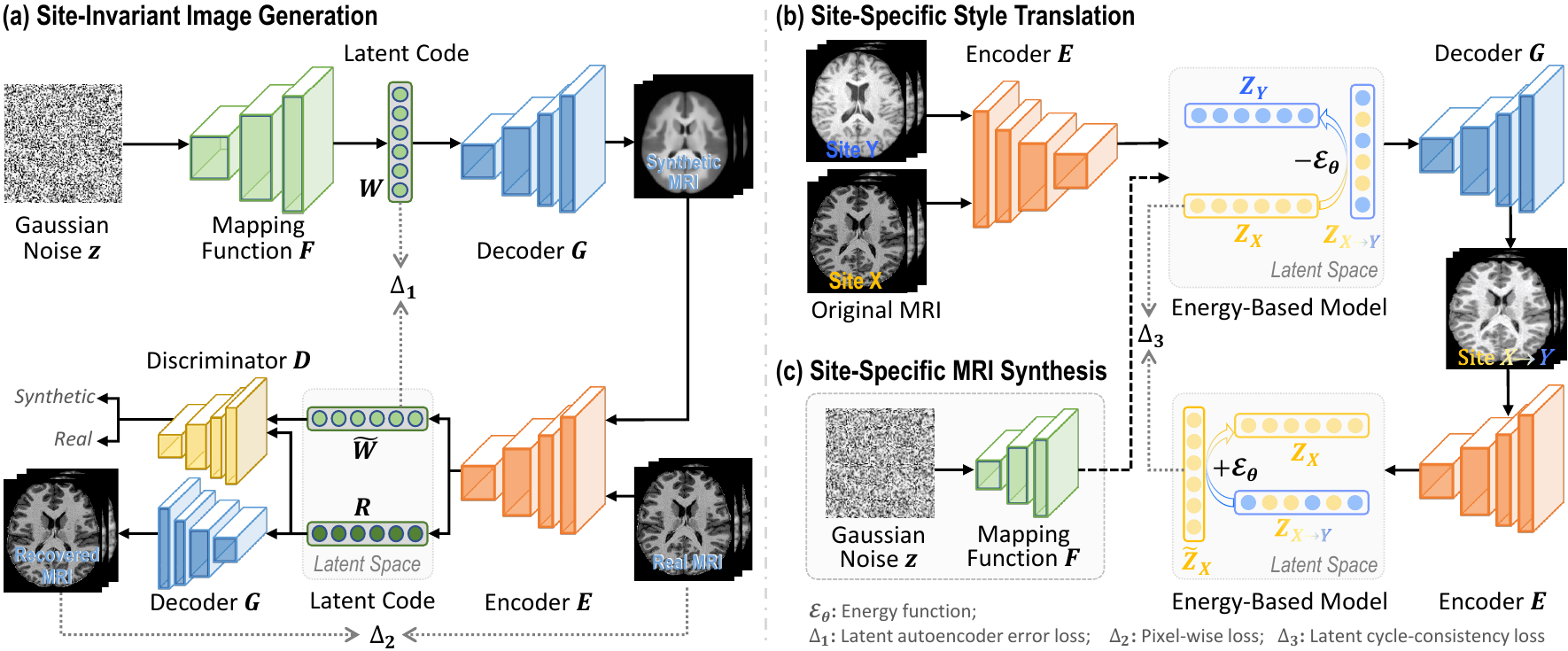}
\caption{The proposed disentangled latent energy-based style translation (DLEST) framework for %image-level 
MRI harmonization at the image level, consisting of (a) \emph{site-invariant image generation} that encodes images into low-dimensional
latent space and reconstruct images based on latent codes, (b) \emph{site-specific style translation} that facilitates implicit style translation within the latent space, and (c) \emph{site-specific MRI synthesis} that generates diverse synthetic MRIs with a given target site style.
%These two modules are disentangled, 
}
\label{fig_pipeline}
\end{figure*}
%%%%%%%%%end%%%%
\section{Introduction}
% Recent advances in machine learning (ML) have enabled the development of computer-aided diagnostic tools to improve the clinical diagnosis of neurodegenerative diseases. Previous studies \citep{Zhang_Wang_Zhou_Yuan_Shen_ADI_2011,Daliri_2012,Mohamed_2019} have shown that such ML methods can be effectively applied in dementia diagnosis and may achieve comparable if not superior diagnostic performance than experienced physicians \citep{Etminani_2022}. However, most of these ML models remain to be useful only in research settings, far away from being applied clinically. One reason for the lack of clinical translation is that these models usually suffer from the generalizability problem, meaning that the model may have significantly suboptimal performance when it is applied to the data with a different distribution than it was trained on. 
Structural magnetic resonance imaging (MRI) has found extensive application in both clinical diagnostics and research studies, primarily due to its ability to offer high spatial resolution in a non-invasive manner~\citep{evans2020non,albayram2022non,varela2017novel}.
%Many large-scale data initiatives and neuroimaging studies have trained machine learning (ML) models on multi-source MRI datasets, pooled from different acquisition sites, to increase the models' statistical power and generalizability. 
Large-scale data initiatives and neuroimaging studies often utilize brain MRI data pooled from various acquisition sites/settings to enhance subject cohort and boost the statistical power of learning models~\citep{Goal_specific,tofts2011multicentre,schnack2010mapping}. 
% {\color{red}However, achieving model convergence on these multi-site datasets can be challenging~\citep{Glocker_scanner_effect, Wachinger_2021}, and generalizability to independent data may be limited due to the sensitivity of MRI to non-biological variations, known as the \emph{site effects}.} 
However, models directly trained on these multi-site datasets often encounter difficulties in achieving convergence~\citep{Glocker_scanner_effect, sun2021multi} 
and in generalizing to new data, due to the sensitivity of MRI to non-biological variations, known as \emph{site effects}. 
The site effect can be attributed to a variety of factors, including differences in field strengths, scanner vendors, scanning protocols, and hardware upgrade and maintenance.~\citep{Glocker_scanner_effect,wachinger2021detect,helmer2016multi}.
While multi-site studies, such as Alzheimer's Disease Neuroimaging Initiative~\citep{ADNI}, aim to mitigate site effects during data acquisition through unified scanning protocols, certain variations are inevitable, like those caused by scanner maintenance or software upgrades. 
Additionally, many studies rely on the retrospective pooling of multi-site data, when there was no coordination among different sites during image acquisition~\citep{OpenBHB}. 
One possible solution to address such a challenge is to retrospectively harmonize the imaging data as a preprocessing step~\citep{ImUnity_2023,CALAMITI_2021}.
% {\color{red}
% One possible solution to address such challenge is to retrospectively harmonize the imaging data before model training.} 

\if false
However, studies \citep{Glocker_scanner_effect, Wachinger_2021} have shown that it is usually challenging for a model to converge on the multi-source dataset, as the MRI data is sensitive to non-biological variations caused by \emph{site effects} such as differences in scanner vendors, field strengths, scanning sequences, and software versions. 
Existing multi-site studies usually suffer from poor model generalizability, as the MRI data is sensitive to non-biological variations caused by \emph{site effects} such as differences in scanner vendors, field strengths, scanning sequences, and software versions. 
%The site effects may include differences in scanner vendors, field strengths, scanning sequences, and software versions. 
Although many cross-site studies have provisioned unified scanning protocols to mitigate site effects during data acquisition, some variations are still inevitable, such as changes caused by scanner maintenance and upgrades. 
Moreover, many studies rely on the retrospective pooling of multi-site data, in which there was no coordination among different sites during the image acquisition time.
Hence, one possible solution to improve the result of these multi-site studies is to harmonize the image data, as a preprocessing step, prior to model training.
\fi

\if false
Previous studies have attempted to harmonize the MRI data on a feature level. This approach usually requires the pre-extraction of image, biological, or radiomic features from a set of predefined regions of interest (ROI). These features are then fed into statistical models, such as ComBat \citep{ComBat} and ComBat-GAM\citep{ComBat_GAM}, which estimate and subtract the site effects using the empirical Bayes method. However, the feature-level harmonization approach is often limited by its dependency on large data size and the quality of feature extraction and is known to suffer from poor generalization on unseen data.
\fi

Existing methods for MRI harmonization can generally be classified into two categories: \emph{feature-level} approaches and \emph{image-level} approaches. 
Feature-level techniques, such as ComBat~\citep{ComBat} and ComBat-GAM~\citep{ComBat_GAM}, typically utilize empirical Bayes models and pre-extracted biological or radiomic features from a set of pre-defined regions-of-interest (ROIs). 
Nonetheless, the effectiveness of these methods is largely contingent on the size of datasets and the quality of feature extraction, which restricts their versatility across applications that employ diverse MRI features~\citep{ImUnity_2023,Goal_specific}.
On the other hand, image-level harmonization is designed for a wider range of downstream applications. 
It is not reliant on pre-defined MRI features, which is the primary objective of this paper.

% {\color{red}Recent research has highlighted the considerable potential of deep generative models in the realm of image-level harmonization~\citep{CycleGAN,StarGANv2,StyleGAN,CALAMITI_2021,ImUnity_2023}.} %%
%{\color{red} 
For image-level harmonization, many deep generative models have been employed in recent studies~\citep{CycleGAN,StarGANv2,StyleGAN,CALAMITI_2021,ImUnity_2023}. % have %effectively employed deep generative models for image-level harmonization.}
These methods, rather than harmonizing a set of pre-extracted MRI features, approach the harmonization problem as a pixel-to-pixel translation task at the image level.
The objective is to alter the style of a source MRI to match a target MRI while preserving the original content. 
In this context, the \emph{style information} refers to high-level features like intensity difference, contrast variations, and signal-to-noise ratio, while \emph{content information} pertains to low-level image features such as anatomical structures, contours, and edges. 
Several state-of-the-art (SOTA) methods utilize generative adversarial networks (GANs) for image-level MRI harmonization.  
For instance, CycleGAN~\citep{CycleGAN} imposes cycle-consistency constraints in its loss function to enforce content preservation during unpaired style translation. 
StarGAN~\citep{StarGANv2} employs separate encoders to learn site-specific style codes and injects them into images via adaptive instance normalization. 
% CALAMITI~\citep{CALAMITI_2021} learns a globally disentangled latent space containing both anatomical and contrast information to achieve harmonization.
ImUnity~\citep{ImUnity_2023} incorporates additional encoder networks as extra constraints for learning %for the 
latent representations to %unlearn 
remove 
the site bias and retain clinical information. 
In addition to GAN-based models, some recent works utilize encoder-decoder models to learn disentangled latent representations for anatomical structure and contrast to perform %to perform %achieve similar 
image-level MRI harmonization \citep{zuo2022disentangling,dewey2020disentangled}.
However, existing image-level methods usually utilize deep networks to separately learn style and content encoding in image space, leading to significantly increased training time and computational cost due to the optimization of numerous network parameters~\citep{gulrajani2017improved}. 
Furthermore, many approaches need to retrain their models entirely when applied to unseen MRI data, therefore limiting their practical generalizability~\citep{ImUnity_2023}.
In this paper, we propose a novel disentangled latent energy-based style translation (\textbf{DLEST}) framework for unpaired structural MRI harmonization at the image level. 
As illustrated in Fig.~\ref{fig_pipeline}, the DLEST includes (a) a \emph{site-invariant image generation} module, using a latent autoencoder for encoding and generating MRIs from a lower-dimensional latent space, (b) a \emph{site-specific style translation} module, using an energy-based model for implicit style translation through latent space, and (c) a \emph{site-specific MRI synthesis} module for generating synthetic MRIs with given site-specific styles. 
%{\color{blue}
Through disentangling image generation and style translation in latent space, the DLEST can achieve efficient and generalizable %multi-source to target 
MRI harmonization. It can also be extended to synthesize MRIs with site-specific styles and diverse %faithful 
anatomical details, %and site-specific styles, 
without extra model training. 
%} 
%extra training cost}. 
% The DLEST consists of two components: (1) A \emph{site-invariant image generation} module, which consists of a latent autoencoder that encodes images into lower-dimensional latent space and generates images from latent code; (2) A \emph{site-specific style-translation} module, which comprises an Energy-based Model \citep{LeCun_2006} (EBM) operates in between the autoencoder, enabling implicit style translation from a source domain to a target domain through latent space.
% The DLEST performs image generation and style translation in low-dimensional latent spaces to reduce the time and computational cost of optimization. 
% The disentanglement of image generation and style translation greatly improves its generalization power to independent data. % without model retraining. 
% We evaluate the DLEST on 4,092 T1-weighted MRIs 
% from OpenBHB~\citep{OpenBHB} and SRPBS~\citep{SRPBS_TS}, 
%  in three tasks: (1) site classification, (2) histogram comparison, and (3) brain tissue segmentation, with extensive 
% with results suggesting its superiority over several state-of-the-art (SOTA) methods. %image-level harmonization methods. 
% To our knowledge, this is one of the first efforts to disentangle image generation and style translation for MRI harmonization. 

An initial version of our work has been introduced in MLMI 2023~\citep{wu2023structural}. 
This journal paper introduces several advancements, including 
1) extension of our framework for site-specific MRI synthesis, with qualitative and quantitative evaluations, 
2) a thorough literature review on image-level harmonization methods and the application of energy-based models for generation tasks, 
3) adding comparison with a recent state-of-the-art method, 
%3) extension of our framework for site-specific MRI synthesis, 
4) experimentation with specific loss functions for content and style regularization, 
5) adding comparative visualization of brain tissue segmentation results, 
6) including experimental analysis on the influence of target site selection, and 
7) adding quantitative comparison of computational cost across all learning methods. 

The key contributions of this work are outlined as follows.
% Major contributions

\vspace{-4pt}
\begin{itemize}
\item We propose a novel disentangled latent energy-based style translation framework for unpaired %multi-site 
    MRI harmonization at the image level, achieving more computationally efficient style translation in low-dimensional latent space. %and can generalize to new data without retraining.
\vspace{-4pt}
    \item Our framework can generalize to new data without model retraining since the image generation module can directly encode and reconstruct MRIs without site specification, while the style translation module can be fine-tuned during the testing stage with minimal computational cost.
    Once trained for harmonization, our method can synthesize MRIs with the style of a specified site. 
\vspace{-4pt}
    \item  We incorporate an energy-based model to capture the latent data distribution of target MRI scans for harmonization. 
    %shifting the focus away from mapping the source to a single reference
    This enables our model to more effectively account for intra-site variations within the target domain and avoids pre-defining a reference image for harmonization.
\vspace{-4pt}
    \item Our framework is rigorously evaluated on two public datasets with T1-weighted MRIs across four tasks, with 
    % , (1) histogram and clustering comparison, (2) acquisition site classification, (3) brain tissue segmentation, and (4) site-specific synthetic MRI generation. 
    qualitative and quantitative results underscoring the superiority of our DLEST over %two traditional and three 
    SOTA methods. % while reducing computational costs significantly.
%    }
\end{itemize}

\if false
our proposed DLEST framework can achieve:
(1) Efficient translation of multiple source sites to the site-specific styles simultaneously;
(2) Extendable translation to new sites without retraining the network;
(3) Generalizable synthetic MRI generation with site-specific styles.
\fi

\if false
% The autoencoder can be first trained on MRI data from any site, in a fully unsupervised manner. The EBM, can then be trained for site-specific style translation at a very low cost due to its lightweight structure. 
The disentanglement property greatly improves its generalization power on unseen data. When a new dataset is added as the source site, DLEST can harmonize it to an existing target site without any retraining. When the new dataset is selected as the target site, the autoencoder can remain unchanged, while only retraining the EBM to translate the style.

% We further employ a content loss and a style loss function to regulate the style translation and ensure content preservation, since the EBM does not explicitly separate the style code from the content code during the translation. 
To our knowledge, this is among the first attempts to disentangle the image generation and style translation process for MRI harmonization. We train our model on T1-weighted MRIs from the OpenBHB dataset and validate it on the independent SRPBS dataset in three tasks: (1) acquisition site classification, (2) Histogram comparison, and (3) brain tissue segmentation. Qualitative and quantitative results demonstrate our proposed DLEST framework outperforms several state-of-the-art image-level harmonization methods.
\fi

\section{Related Work}
\subsection{Image-Level MRI Harmonization}
% Recent studies have leveraged the concept of image-to-image translation from the natural imaging domain to address medical data harmonization issues at the image level using generative adversarial network (GAN) models.
Inspired by the image-to-image translation task in the natural imaging domain, recent studies have employed %adopted 
the generative adversarial network (GAN) models to tackle medical data harmonization problems on the image level~\citep{CycleGAN,StarGANv2,StyleGAN}. 
% These methods formulate the model training as an adversarial game between the generator network and the discriminator network: The generator learns to generate synthetic images that resemble the distribution of the real dataset, while the discriminator learns to differentiate the synthetic images from the real images. 
These methods engage the generator and discriminator networks in an adversarial game, where the generator creates synthetic images resembling the real dataset distribution, and the discriminator differentiates between synthetic and real images~\citep{CycleGAN}. 
%These GAN-based methods have improved upon one another over recent years. 
For instance, CycleGAN introduces a cycle-consistency constraint in its loss function for unpaired image translation and content (anatomical structure) preservation~\citep{CycleGAN}. 
Style-encoding GAN~\citep{StyleGAN}, inspired by StarGAN-V2~\citep{StarGANv2}, further separates the content and style encoding in the latent space, allowing the site-specific style code to be learned and injected into images via adaptive instance normalization.
% Attention-guided GAN~\citep{AttentionGAN} incorporates an attention mechanism to enhance the style translation in high-level semantic image regions while suppressing the changes in low-level, static regions. 
ImUnity~\citep{ImUnity_2023} modifies the GAN structure by adding a site/scanner unlearning module to encourage the encoder to learn domain-invariant latent representations. 
% Additionally, it incorporates a biological preservation module to improve anatomical structural preservation. 
These advancements have contributed to the continual improvement of GAN-based harmonization methods.

In addition to GAN-based models, recent studies have introduced an alternative approach that employs encoder-decoder networks to disentangle anatomical and contrast information in latent space. 
For instance, CALAMITI~\citep{CALAMITI_2021} first uses T1- and T2-weighted (T1/T2-w) MRI pairs to learn global latent codes containing anatomical and contrast information, and then disentangles %the latent codes to represent 
 style and content latent codes via separate encoders and decoders.  
%{\color{red}CALAMITI~\citep{CALAMITI_2021}, for instance, learns a global latent space containing both anatomical and contrast information, it then disentangles the latent information via a style encoder, a contrast encoder, and a style discriminator using T1-weighted and T2-weighted MRI pairs.} 
Similarly, Dewey~\etal~\citep{dewey2020disentangled} leverage T1-w and T2-w image pairs to attain a disentangled latent space, comprising high-resolution anatomical and low-dimensional contrast components via a Randomization block. This block allows 
% randomized anatomical and contrasts latent code combinations, 
generating MRIs with identical anatomical structures but varying contrast. Zuo~\etal~\citep{zuo2022disentangling} enhance this approach by ablating the need for paired MRI sequences. Instead, they employ 2D slices from axial and coronal views of the same MRI to provide the same contrast but different anatomical information. 

These existing methods usually %share several common limitations. 
%For example, some methods 
involve simultaneous training of multiple encoders, decoders, or sub-networks for separated style and content encoding, resulting in increased computational and time costs for model training. 
Besides, some methods have limited generalizability on new MRI as their models need to be retrained entirely on new datasets.
Moreover, certain existing studies necessitate paired MRI scans for model
training, like traveling subjects or multiple MRI sequences,
% (T1-weighted and T2-weighted)
which are often unavailable in practice. %retrospective studies.  
%Besides, many of these approaches suffer from limited image generation power due to mode collapse. 
In this work, we aim to enhance the efficiency of image-level harmonization and the model's generalizability to new data, while also expanding its capacity for site-specific image synthesis.

\if false
{\color{red}
CycleGAN 
% and Attention-guided GAN 
function primarily for one-to-one MRI harmonization, which means they need to be fine-tuned for each source-target site pair or completely retrained when applied to new data~\citep{ImUnity_2023}. 
StarGAN treats each scan as a unique domain with its own style and harmonizes the source images to a single target scan~\citep{StyleGAN}. This approach may fail to capture the global distribution of the target site and may ignore the intra-site variations. 
Moreover, certain existing studies necessitate paired MRI scans during
training, like traveling subjects or multiple MR sequences,
% (T1-weighted and T2-weighted)
which are often unavailable in retrospective studies. 
Additionally, some methods involve simultaneous training of multiple encoders, decoders, or sub-networks for separated style and content encoding, resulting in increased computation and time cost for model training. 
Besides, many of these approaches suffer from limited image generation power due to mode collapse. Thus, our proposed framework aims to enhance both the efficiency of image-level harmonization and the model's ability to generalize to independent data, while also expanding its capacity for site-specific image synthesis.
}
\fi 
% These GAN-based methods have improved upon one another over recent years. Conditional GAN \citep{Isola_Zhu_Zhou_Efros_2017} uses paired, supervised training images to allow the generated synthetic images to be conditioned on the specific inputs. CycleGAN \citep{Zhu_Park_Isola_Efros_2017} imposes the cycle-consistency constraints in its loss function to ensure content preservation during unpaired style translation. StarGAN \citep{Choi_Uh_Yoo_Ha_2019} further separated the style and content in the latent space, allowing the site-specific style code to be learned and injected into images using adaptive instance normalization. The attention mechanism \citep{attention_GAN_AD} has also been applied to the model to improve the style translation in high-level semantic regions while suppressing the changes in low-level, static regions. A common limitation of the GAN-based methods is the high cost and difficulty of converging during the model training because the models consist of multiple networks that need to be trained simultaneously. The cycle-consistency loss calculated in the ambient space also increased the training cost.

\subsection{Energy-Based Models for Image Generation}
%Generative models are a type of probabilistic model, which models a joint probability of the observed data and the target data. 
Generative models are a category of probabilistic models that represent the joint probability distribution of both observed and target data. 
For the image generation task, a model observes image samples or their latent representations and learns to capture the true data distribution to generate new images that resemble the observed samples~\citep {Pang_2020_Latent_EBM,du2019implicit}.
Probabilistic models employ statistical inference methods, such as maximum likelihood estimation (MLE), to estimate parameters that maximize the likelihood of the observed data under an assumed distribution \citep{ng_2001}. 
However, these models need to be correctly normalized. 
This process can sometimes involve the evaluation of intractable integrals over all possible variable configurations, which can be computationally challenging. 
Therefore, LeCun \etal\citep{LeCun_2006} use energy-based learning as an alternative framework 
% to probabilistic approaches 
for both discriminative and generative tasks. 
Energy-based models (EBMs) capture data dependencies by associating a scalar energy value (a measure of compatibility) to each configuration of variables \citep{LeCun_2006}. 
Unlike probabilistic models, EBMs do not require normalization. 
For generative tasks, the training phase of EBMs involves learning an appropriate energy function that assigns low energy values to observed data within the real data distribution and high energy values otherwise, during which data is typically sampled via Markov Chain Monte Carlo (MCMC) sampling methods. 
During inference, the EBMs generate data by exploring the energy surface to sample configurations that minimize the energy function.

Over the years, many studies have employed EBMs for image generation and improved on various aspects. Du \etal \citep{du2019implicit} propose to utilize Langevin Dynamics, a gradient-based MCMC method, for more efficient sampling and mixing on high-dimensional image domains. 
Pang \etal \citep{Pang_2020_Latent_EBM} further improves the sampling efficiency and the MCMC mixing by combining an EBM with a top-down network (decoder) in the low-dimensional latent space. The EBM serves to refine the simple isotropic Gaussian prior to be closer to the data distribution, leading to improved quality during image generation. Xiao \etal \citep{Xiao_2021_VAEBM} propose a symbiosis of Variational Autoencoder (VAE) and EBM to leverage the advantages of both. The VAE captures the overall model structure of the data distribution and provides an improved latent initialization of the MCMC sampling, which is used in EBM. The EBM, on the other hand, can exclude non-data-like samples and refine the latent representation for VAE to decode. 
A recent work \citep{zhao2021unpaired} utilizes this symbiotic combination of EBM and autoencoder for the task of male-to-female face translation. In this context, the EBM learns the latent distribution of a target image domain and translates the source latent representation towards this target distribution. The underlying assumption is that two image domains share certain common latent representations. The EBM is then able to implicitly translate latent codes that encapsulate domain differences while preserving shared latent codes that signify domain similarities. 
Inspired by the success of EBMs in natural image-to-image translation, our method adopts a similar assumption. We posit that MRIs from different domains share some common latent representations that convey content (anatomical) information. 
We propose training an EBM to translate the site-specific style information from one domain to another within a low-dimensional latent space. 
%However, compared to natural images, medical images like MRI contain much more nuanced constraints that need to be further considered and regularized. 
%We shall elaborate in detail in the subsequent sections.

\section{Materials and Methodology}

\subsection{Materials and Image Preprocessing}
\subsubsection{Datasets}
\if false
We evaluate the proposed DLEST harmonization framework on two public datasets: (1) Open Big Healthy Brains (OpenBHB) \citep{OpenBHB}, (2) Strategic Research Program for Brain Science (SRPBS) \citep{SRPBS_TS}. 
The OpenBHB is a large-scale brain MRI dataset, containing T1-weighted MRIs from healthy controls, gathered from $\geq 60$ centers worldwide. 
It has been split into a training set (with 3,227 subjects from 58 acquisition sites/settings) and a validation set (with 757 subjects). 
The SRPBS contains 108 T1-weighted MRIs from 9 healthy traveling subjects with 12 scanners/settings. 
\fi
%We evaluate the proposed DLEST harmonization framework 
Our study utilizes two public datasets, including (1) Open Big Healthy Brains (OpenBHB) \citep{OpenBHB} and (2) Strategic Research Program for Brain Science (SRPBS) \citep{SRPBS_TS}. 
Specifically, the OpenBHB, a large-scale brain MRI dataset, comprises 3,984 T1-weighted MRIs from healthy subjects acquired from over 58 centers globally. 
The SRPBS dataset contains 108 T1-weighted structural MRIs acquired from 9 healthy traveling subjects across 12 scanners/settings. 
In this work, MR scans of 9 subjects from 11 acquisition sites/settings in SRPBS are used, while one duplicated site (YC2) is removed. 
Further details %on these two datasets 
can be found online\footnote{\hyperlink{}{https://baobablab.github.io/bhb/}}$^,$\footnote{\hyperlink{}{https://www.synapse.org/\#!Synapse:syn22317076/wiki/605026}}.
%Further details are available in the Supplementary Materials.

% Two public datasets are used: (1) Open Big Healthy Brains (OpenBHB) \citep{OpenBHB}, (2) Strategic Research Program for Brain Science (SRPBS) \citep{SRPBS_TS}. 
% The OpenBHB is a large-scale brain MRI dataset, containing T1-weighted MRIs from healthy controls, gathered from $\geq 60$ centers worldwide. 
% It has been split into a training set (with 3,227 subjects from 58 acquisition sites/settings) and a validation set (with 757 subjects). 
% The SRPBS contains 108 T1-weighted MRIs from 9 healthy traveling subjects with 12 scanners/settings. 
% More details can be found in the \emph{Supplementary Materials}. 

\subsubsection{Data Preprocessing}
For these two datasets, each MRI scan undergoes minimal preprocessing using FSL ANAT pipeline~\citep{FSL}, which encompasses bias field correction, brain extraction, and 9 degrees of freedom linear registration to the MNI-152 template with $1\,mm$ isotropic resolution. 
Brain tissue segmentation is performed on SRPBS images using FSL FAST pipeline, with each brain segmented %dividing them 
into white matter (WM), gray matter (GM), and cerebrospinal fluid (CSF). 
For each MRI from OpenBHB, we select 10 slices in axial view from the center of each MRI.
We utilize the original training and test data split specified in~\cite{OpenBHB}, which eventually results in $32,270$ 2D MRI slices for training and $7,570$ MRI slices for validation.
For SRPBS with limited traveling subject MRIs, we choose 14 slices from each volume, resulting in $1,386$ 2D MRI slices.
Additionally, all these 2D slices are zero-padded to have the size of $256\times256$.

\subsection{Proposed Methodology}
%\subsection{Problem Formulation}
\subsubsection{Problem Formulation}
%{\textbf{Problem Formulation}}. 
In this work, we reformulate image-level MRI harmonization as a \emph{domain-level translation} task, which is different from previous studies that treat each MRI scan as a unique domain and concentrate on individual source-to-reference image translation~\citep{StyleGAN,ImUnity_2023}. 
%Contrary to previous studies \citep{StyleGAN,ImUnity_2023} that treat each MRI scan as a unique domain and concentrate on individual source-to-reference image translation, our framework reformulates the problem as a domain translation task. 
In the context of multi-site MRI studies, we consider %MRI scans from 
a single site or scanning setting for MRI acquisition as a unique domain. 
%The shift from individual image-level translation to domain-level translation 
This is justified by numerous studies demonstrating that MRIs scanned in the same site under the same setting can exhibit intra-site variations
that result in different styles within one domain~\citep{deoni2008standardized,styner2002multisite}. 
Consequently, image-level translation using only one reference image may not account for these variations, failing to capture true data distributions of a site/domain.
%In this work, we propose to use all MRIs from the same site as the target domain to better encapsulate its style characteristics.
In this work, we propose to employ all MRIs from a target site as reference images, aiming to better encapsulate the stylistic features of target data. 
% For the two datasets used in this work, we first compute the mean peak signal-to-noise ratio (PSNR) for each domain by averaging the PSNR values obtained from every unique pair of images within that same domain.
% We then designate the domain with minimal intra-site variations (highest PSNR) as the \emph{target domain}, with the remaining site(s) as the \emph{source domain(s)}.
%Overall, our harmonization framework 
%In this work, we aim to translate imaging style (\ie, non-biological variation) of MRIs within each source domain $X$ to align with the style of MRIs in a target domain $Y$. 
Overall, we aim to translate MRIs within a source domain $X$ to have the imaging style (\ie, non-biological variation) of MRIs in a target domain $Y$.

Specifically, our goal is to train a latent autoencoder that can map a source and a target MRI scan to their respective latent codes, denoted as $\bm{E}$: $\{\x\in X, \y\in Y\}$$\to$$ \{Z_x, Z_y\}$. 
In this context, a latent code $Z=\{S, C\}$ represents an MRI scan in low-dimensional space by implicitly encapsulating both a \emph{style code} $S$ and a \emph{content code} $C$. %, denoted as $Z=\{S, C\}$.
The source style code $S_x$ is then implicitly translated to become closer to the style code $S_y$ of the target domain, while the content code $C_x$ is preserved. 
This style translation process can be formulated through a mapping $\bm{T}$: $Z_x$$=$$\{S_x,C_x\}$$\to$$Z_{x\to y}$$=$$\{S_y,C_x\}$, where $Z_{x\to y}$ represents the latent code of an MRI after translation.
% By translating the style code $S_x$ of the source domain to the target domain and also maintaining its content code , we can formulate the style translation as a mapping $\bm{T}$: $Z_x$$=$$\{S_x,C_x\}$$\to$$Z_{x\to y}$$=$$\{S_y,C_x\}$, where $Z_{x\to y}$ is the latent code of the source image in target domain. 
Subsequently, using the same autoencoder, we learn another mapping $\bm{G}$: $ Z_{x\to y}$$\to$$\Tilde \y$. 
This latent-to-image space mapping decodes the translated source latent code $Z_{x\to y}$ to generate an image $\Tilde{\y}$ in the target domain. 
This image, preserving the content of $\x$ and adopting the style of $\y$, serves as an effective translation from the source to the target domain.
% Then, we propose to learn another mapping $\bm{G}$: $ Z_{x\to y}$$\to$$\Tilde \y$ via the autoencoder to decode the translated latent code back to the image space, resulting in an image $\Tilde{\y}$ in the target domain with the style of $\y$ and the content of $\x$. 
% In this way, the style of source images can be translated into the target domain while keeping their content. 
To achieve this, as depicted in Fig.~\ref{fig_pipeline}, our DLEST consists of three disentangled modules: \emph{site-invariant image generation}, \emph{site-specific style translation}, and \emph{site-specific MRI synthesis}. 
The specifics of these modules are introduced as follows. %discussed in the subsequent sections.

%%%%% New Section  %%%
\subsubsection{Site-Invariant Image Generation (SIG)}
As shown in Fig.~\ref{fig_pipeline}~(a), the SIG is designed to encode MRI scans into a lower-dimensional latent space and subsequently decode reconstructed MRIs from these latent codes. 
Importantly, this process is trained
on the OpenBHB training set,
\emph{independently of any site specifications}, and thus, does not require any site labels or paired images. 
%autoencoder E and G do not require further training, even when applied to unseen dataset. 
Specifically, the SIG module takes as input a real MRI $\x\sim \mathbb{P}_{Data}$ and a random noisy image $z$ drawn from a Gaussian distribution $\mathcal{N}(0,1)$, where $\mathbb{P}_{Data}$ denotes the real data distribution. 
A network $\bm F$ learns a deterministic mapping to convert $z$ into a latent code~\citep{ALAE}, represented as $\mathcal{W}={F}(z)$. 
Following this, we further design a stochastic decoding network $\bm{G}$ to generate a synthetic MR image based on the latent code $\mathcal{W}$ and an independent noise $\eta\sim \mathcal{N}(0,1)$, represented as $\Tilde{\x}={G}(\mathcal{W},\eta)$. 
%Following this, a stochastic decoding network $\bm{G}$ employs the latent code $\mathcal{W}$ and an independent noise $\eta\sim N(0,1)$ to generate a synthetic MRI scan, represented as as $\Tilde{\x}={G}(\mathcal{W},\eta)$. 
% Specifically, the Gaussian noise    
% A deterministic mapping network  is used to map $z$ into a latent code , followed by a stochastic decoder  that takes the latent code $\mathcal{W}$ with an   
A network $\bm{E}$ is then trained to encode the synthetic image $\Tilde{\x}$ back into the latent space, resulting in a reconstructed latent code $\Tilde{\mathcal{W}}$. This entire process can be symbolized  as $\Tilde{\mathcal{W}}=E(\Tilde{\x})=E\circ G \circ F(z)$, where the symbol $\circ$ denotes function composition. 
The encoder $\bm{E}$ also can transform a real image $\x$ into its corresponding latent code, denoted as $\mathcal{R}=E(\x)$.
Lastly, a latent discriminator, denoted as $\bm{D}$, receives a latent code and produces a probability value that signifies whether the latent code is derived from a synthetic or real MRI. The networks $\bm F$, $\bm G$, $\bm E$, and $\bm D$ are trained using an adversarial approach, enabling the SIG module to generate latent codes that closely resemble those derived from real MRIs. 
The latent code $\mathcal{R}$, derived from the real image, is also decoded back to a reconstructed image using the decoder $\bm{G}$, aiming to preserve the content (\eg, anatomical structures) of the real MRI. %, including its anatomical structures.
% Finally, a discriminator $\bm{D}$ takes a latent code and outputs a scalar value, indicating if that latent code belongs to a real or synthetic image. 
% We also decode the latent code of the real image back to a recovered image via a decoder $\bm{G}$, aiming to preserve the content (\eg, anatomic structure) of the real image. 

%\emph{\textbf{1) Latent Autoencoder Error Loss.}} 
To uphold the symmetry of the encoder and decoder in latent space, we aim to ensure that $\mathcal{W}=\Tilde{\mathcal{W}}$, %which can  
%This can be 
accomplished by minimizing the following \textbf{\emph{latent autoencoder error loss}}:
\begin{equation}
\label{eq:1}
%\small
\mathcal{L}_{lae}^{E,G} = \Delta_1(F\lVert E\circ G\circ F)= \mathbb{E}_{z\sim \mathbb{P}_z}[\lVert F(z)-E\circ G\circ F(z) \rVert_2^2]
\end{equation}
Through this process, 
%By imposing this error loss, 
the network pair $\bm E$ and $\bm G$ can be interpreted as a \emph{latent autoencoder} that encodes and reconstructs the latent code $\mathcal{W}$. This differs from conventional autoencoders, which typically operate in high-dimensional image space~\citep{VAE}. 
The generation of images via autoencoding low-dimensional latent codes enhances computational efficiency.

To maintain the anatomical structure preservation of the latent autoencoder $\bm E$ and $\bm G$, a \textbf{\emph{pixel-wise content loss}} is introduced to compute the difference between the original real image $\x$ and its reconstructed counterpart $\Tilde{\x}$, as follows: 
\begin{equation}
\label{eq:2}
%\small
    \mathcal{L}_{pix}^{E,G}= \Delta_2(\x\lVert\Tilde{\x}) = \mathbb{E}_{x\sim \mathbb{P}_{Data}}[\lVert \x - G\circ E(\x)\rVert_1]
\end{equation}
which motivates the model to minimize content change during MRI reconstruction. 
To generate faithful synthetic MRIs, we propose to jointly optimize $\bm F$, $\bm G$, $\bm E$, and $\bm D$ through an \textbf{\emph{adversarial loss}}, which is formulated as:
\begin{equation}
\label{eq:3}
%\small
\mathcal{L}_{adv}^{E,D}=\Phi(D\circ E\circ G\circ F(z)) + \Phi(- D\circ E(\x)) + R_{reg}
\end{equation}
\begin{equation}
\label{eq:4}
%\small
\mathcal{L}_{adv}^{F,G}=\Phi(-D\circ E\circ G\circ F(z))
\end{equation}
where $\Phi$ is the $softplus$ function defined as $\Phi(t)=\log(1+\exp(t))$, $R_{reg}$ is a zero-centered gradient penalty term defined as $\frac{\gamma}{2}\mathbb{E}_{x\sim \mathbb{P}_{Data}}[\lVert \nabla D\circ E(\x)\rVert^2]$~\citep{ALAE}. 

\if false
Specifically, a deterministic mapping network $\bm F$ first takes a random Gaussian noise $z$ and maps it into a latent code $\mathcal{W}=\bm{F}(z)$. 
A stochastic generator $\bm{G}$ then takes the latent code $\mathcal{W}$ with an independent noise $\eta\sim N(0,1)$ and generates a synthetic image $\Tilde{\x}=\bm{G}(\mathcal{W},\eta)$. 
An encoding network $\bm{E}$ takes the generated synthetic image $\Tilde{\x}$ and encodes it back to latent code $\Tilde{\mathcal{W}}=\bm{E}(\Tilde{\x})$. 
Similarly, the encoder $\bm{E}$ can encode a real image $\x\sim p_{Data}$ into its latent code $\mathcal{R}=\bm{E}(\x)$. 
Finally, a discriminator $\bm{D}$ takes a latent code and outputs a scalar value, indicating if that latent code belongs to a real or synthetic image.
% The ALAE consists of a deterministic mapping network $\bm F$, a pair of encoder-decoder $\bm E, \bm G$, respectively, and a discriminator network $\bm D$, see Fig. \ref{fig_pipeline}. $\bm{F}$ takes a random Gaussian noise $z$ and maps it into a latent code $\mathcal{W}=\bm{F}(z)$. The stochastic generator $\bm{G}$ then takes the latent code $\mathcal{W}$ with an independent noise $\eta\sim N(0,1)$ and generates a synthetic image $\Tilde{x}=\bm{G}(\mathcal{W},\eta)$. The encoder $\bm{E}$ takes the synthetic image generated by $\bm{F}$ and $\bm{G}$ and encodes it back to latent code $\Tilde{\mathcal{W}}=\bm{E}(\Tilde{x})$. If we replace the synthetic image $\Tilde{x}$ with a real image $x\sim p_{Data}$, we get the encoded latent code $\mathcal{R}=\bm{E}(x)$ for the real image. Finally, the discriminator $\bm{D}$ takes a latent code and outputs a scalar value, indicating if the latent code belongs to a real or synthetic image.

\emph{\textbf{1) Adversarial Loss.}} Network $\bm F$,$\bm G$, $\bm E$, and $\bm D$ are optimized by the adversarial loss:
\begin{equation}
\label{eq:1}
\mathcal{L}_{adv}^{E,D}=\Phi(D\circ E\circ G\circ F(z)) + \Phi(- D\circ E(\x)) + R_1
\end{equation}
\begin{equation}
\label{eq:2}
\mathcal{L}_{adv}^{F,G}=\Phi(-D\circ E\circ G\circ F(z))
\end{equation}
where $\Phi$ is the $softplus$ function, defined as $\Phi(t)=\log(1+\exp(t))$; The $R_1$ is a zero-centered gradient penalty term, defined as $\frac{\gamma}{2}E_{x\sim P_{Data}}[\lVert \nabla D\circ E(\x)\rVert^2]$. 

\emph{\textbf{2) Latent Autoencoder Error Loss.}} To ensure the reciprocity of the autoencoder in latent space, we require that $\mathcal{W}=\Tilde{\mathcal{W}}$, thus formulate the latent autoencoder error loss:
\begin{equation}
\label{eq:3}
\mathcal{L}_{lae}^{E,G} = \Delta(F\lVert E\circ G\circ F)= \mathbb{E}_{z\sim P_z}[\lVert F(z)-E\circ G\circ F(z) \rVert_2^2]
\end{equation}
By imposing this error loss, the pair of networks $\bm E$ and $\bm G$ can be considered as a \emph{latent autoencoder} that autoencode the latent code $\mathcal{W}$, instead of an image.

\emph{\textbf{3) Pixel Loss.}} 
To ensure the content preservation of latent autoencoder $\bm E$ and $\bm G$, we impose the following pixel loss by comparing a real image with its recovered version 
\begin{equation}
\label{eq:4}
    \mathcal{L}_{pix}^{E,G}=\mathbb{E}_{x\sim P_{Data}}[\lVert \x - G\circ E(\x)\rVert_1]
\end{equation}
\fi

%\noindent\textbf{Full Objective.} 
Overall, the final objective functions of SIG are as follows:
\begin{equation}
\label{eq:5}
%\small
%\underset{F,G}
{\operatorname{min}_{\{F,G\}}} \, %\underset{E,D}
{\operatorname{max}_{\{E,D\}}} \, \mathcal{L}_{adv}^{E,D} + \mathcal{L}_{adv}^{F,G}, ~~
%\end{equation}
%\begin{equation}
%\label{eq:6}
%\underset{E,G}
{\operatorname{min}_{\{E,G\}}} \, \mathcal{L}_{lae}^{E,G}+\mathcal{L}_{pix}^{E,G}
\end{equation}

% \noindent\textbf{Training.} The training of ALAE can be completed in a fully unsupervised manner on the entire multi-site dataset without requiring any site labels. Each iteration consists of three optimizing steps: the first and the second steps update network $\bm{E,D}$ and $\bm{F,G}$, respectively, corresponding to Eq. \ref{eq:5}. The third step update the latent space autoencoder $\bm{E,G}$, corresponding to Eq. \ref{eq:6}. Compared to similar GAN models, the overall training cost of ALAE is lower because both objective functions are evaluated in low-dimensional latent space, and no additional cost is incurred for Cycle-consistency constraints. After completing the training, the latent autoencoder $\bm E$ and $\bm G$ can be used in the energy-based style translation model, with their weights frozen.

%%%%%%% -- New Section  --- %%%%%%%%
\subsubsection{Site-Specific Style Translation (SST)}
As depicted in Fig.~\ref{fig_pipeline}~(b), the proposed SST consists of (1) an autoencoder that includes the networks $\bm E$ and $\bm G$ trained in the  SIG module, and (2) 
an energy-based model (EBM) that performs latent-space style translation. 
Given an MRI $\x$ from source site $X$ and an MRI $\y$ from target site $Y$, the encoder $\bm E$ initially transforms them into their respective latent codes, denoted as $Z_x=E(\x)$ and $Z_y=E(\y)$, respectively. 
These latent codes are then input into the EBM, which produces the translated latent code $Z_{x\to y}$. This translated latent code is subsequently decoded by $\bm{G}$ to produce the translated MRI. 
The EBM is also utilized to reverse-translate $Z_{x\to y}$, resulting in a recovered source latent code, denoted as $\Tilde{Z}_x$. This reconstructed latent code is compared with the original $Z_x$ to ensure the latent cycle-consistency, which prevents mode collapse during the unpaired style translation process.
The specifics of the EBM are discussed as follows.
%impose cycle-consistency \citep{CycleGAN} constraint. 

%\noindent\textbf{EBM.} 
The energy-based model, through observing latent code samples, learns to model a distribution that is close to the true data distribution of a target site in latent space, denoted as $\mathbb{P}_\theta(Z_y)\sim \mathbb{P}_{Data}(Z_y)$.
Specifically, we assume that $\mathbb{P}_\theta(Z_y)$ %is assumed to 
follows a Gibbs distribution \citep{du2019implicit,Xiao_2021_VAEBM}, which is defined as $\mathbb{P}_\theta(Z_y)=\exp({-\mathcal{E}_\theta(Z_y)})/Q(\theta)$, 
% \begin{equation}
% \label{eq:7}
% P_\theta(Z_y)=\frac{1}{Q(\theta)}\exp({-\mathcal{E}_\theta(Z_y)})
% \end{equation}
where $\mathcal{E}_\theta(Z_y):\mathbb{R}^D$$\to$$ \mathbb{R}$ is a scalar energy function parameterized by $\theta$. 
This function $\mathcal{E}_\theta(Z_y)$ is trained to produce lower energy values to observed latent code samples within the target latent distribution and higher energy values to other inputs.  
And $Q(\theta)=\int{\exp(-\mathcal{E}_\theta(Z_y))\ dZ_y}$ is the partition function for probability normalization, which is intractable. 
% The EBM can be trained by finding the $\theta$ that maximizes the log-likelihood $L(\theta)=\mathbb{E}_{Z_y\sim P_{Data}}(\log(P_\theta(Z_y)))$. This can be achieved by maximizing the derivative of the negative log-likelihood or minimizing its inverse
% One can optimize EBM by maximizing the log-likelihood or equivalently minimizing the negative log-likelihood using its derivative, formulated as: 
% \begin{equation}
% \label{eq:6}
% %\small
% \footnotesize
% \begin{split}
% \mathcal{L}_{EBM} &=-(\frac{\partial}{\partial\theta}L(\theta)) \\
% &=\mathbb{E}_{Z_y\sim \mathbb{P}_{Data}}[\frac{\partial}{\partial\theta}\mathcal{E}_\theta (Z_y)]-\mathbb{E}_{\Tilde{Z}_y\sim \mathbb{P}_\theta}[\frac{\partial}{\partial\theta}\mathcal{E}_\theta (\Tilde{Z}_y)]
% \end{split}
% \end{equation}
One can optimize an EBM by maximizing the log-likelihood or equivalently minimizing the negative log-likelihood. The gradient of the negative log-likelihood is formulated as:
\begin{equation}
\label{eq:6}
%\footnotesize
\begin{split}
\nabla_\theta \mathcal{L}_{EBM} &= -\frac{\partial}{\partial\theta} L(\theta) \\
&= \mathbb{E}_{Z_y \sim \mathbb{P}_{Data}} \left[ \frac{\partial}{\partial\theta} \mathcal{E}_\theta(Z_y) \right] - \mathbb{E}_{\tilde{Z}_y \sim \mathbb{P}_\theta} \left[ \frac{\partial}{\partial\theta} \mathcal{E}_\theta(\tilde{Z}_y) \right]
\end{split}
\end{equation}

where the first expectation term $\mathbb{E}_{Z_y}$ corresponds to samples $Z_y$, which are drawn from the actual target data distribution $\mathbb{P}_{Data}(Y)$. 
On the other hand, the second expectation term $\mathbb{E}_{\Tilde{Z}_y}$ 
 corresponds to samples $\Tilde{Z}_y$, which are drawn from the modeled distribution $\mathbb{P}_\theta$. 
This requires direct sampling from $\mathbb{P}_\theta$, an intractable model distribution.  
This can be approximated using a Markov chain Monte Carlo (MCMC) sampling method, namely the Stochastic Gradient Langevin Dynamics (SGLD)~\citep{SGLD}, an iterative method defined as follows: 
\begin{equation}
\label{eq:7}
%\small
\Tilde{Z}_y^{t+1}=\Tilde{Z}_y^t-\frac{\eta^t}{2}\frac{\partial}{\partial\Tilde{Z}_y^t}\mathcal{E}_\theta(\Tilde{Z}_y^t)+\sqrt{\eta^t}\epsilon^t,\ \epsilon^t\sim \mathcal{N}(0,I)
\end{equation}
where $t=[0,\cdots,T]$ represents the sampling steps with a step size $\eta$ and $\epsilon$ is a Gaussian noise that is injected to account for data uncertainty and to improve sample convergence~\citep{LeCun_2006}. 
The initial $\Tilde{Z}_y^0$ are latent codes sampled from the source site $X$, denoted as $\Tilde{Z}_y^0=Z_x=E(\x)$. 
The SGLD sampling process involves iteratively updating the initial source latent codes using negative gradients of the energy function $\mathcal{E}_\theta$. 
This sampling process can be interpreted as evolving original source latent codes towards a configuration with lower energy, thereby \emph{aligning them more closely with the true target latent distribution}. 
In contrast to source-to-reference image translation~\citep{StyleGAN,CycleGAN,StarGANv2}, EBM effectively traverses the latent space of the entire target domain, allowing a comprehensive consideration for intra-site variations within the target domain.
After $T$ sampling steps (via SGLD), we can decode $\Tilde{Z}_y^T$ %can be decoded 
through $\bm G$ to generate a translated MRI, denoted as $\Tilde{\y}=G(Z_{x\to y})=G(\Tilde{Z}_y^T)$.

%\noindent\textbf{Latent Content Loss.} 
Considering that the EBM does not differentiate the content code and the style code explicitly during style translation, we further introduce a \textbf{\emph{latent content loss}} into the EBM for regularization measure. This approach is designed to prompt the model to minimally alter the latent code during style translation, thereby preserving the overall content. This loss function measures the difference between the translated latent code $Z_{x\to y}$ and the original source latent code $Z_x$, which is formulated as: % to minimize the  during translation via 
\begin{equation}
\label{eq:8}
    \mathcal{L}_{con}=\mathbb{E}_{Z_x\sim \mathbb{P}_{Data},Z_{x\to y}\sim \mathbb{P}_{\theta}}[\lVert Z_x-Z_{x\to y} \rVert_1]
\end{equation}
Instead of using the latent content loss in Eq.~\eqref{eq:8}, one can also explicitly compute the content loss and the style loss during the latent style translation, which will be discussed in Section~\ref{sec:csloss}.

%\noindent\textbf{Latent Cycle-Consistency Loss.} 
To guarantee the learned translation is bijective and to further avoid mode collapse during unpaired latent style translation, we introduce a \textbf{\emph{latent cycle-consistency loss}}, defined as:
\begin{equation}
\label{eq:9}
    \mathcal{L}_{cyc}=\mathbb{E}_{Z_x\sim \mathbb{P}_{Data}}[\lVert Z_x - \mathbb{P}_{\theta}^{-1}\circ E\circ G\circ \mathbb{P}_{\theta}(Z_x) \rVert_1]
\end{equation}
where $\mathbb{P}_{\theta}^{-1}$ represents the same EBM as $\mathbb{P}_{\theta}$, but it operates with an inverse SGLD sampling process. 
This means that it will update the latent code using positive gradients of the energy function, in contrast to the negative gradients used by $\mathbb{P}_{\theta}$.

With the integration of Eqs.~\eqref{eq:6}, \eqref{eq:8}-\eqref{eq:9}, the full objective function of SST can be written as: 
\begin{equation}
\label{eq:10}
    \mathcal{L}(\theta)=\mathcal{L}_{EBM} + \alpha \mathcal{L}_{con}+\beta \mathcal{L}_{cyc}    
\end{equation}
where $\alpha$ and $\beta$ serve as hyperparameters that determine the relative contributions of the three terms in the equation. %weight of each loss. 
The SST can be trained or fine-tuned for any target style translation with a small sample size for each target site (i.e., $\sim$$100$ MRI
slices in our experiments).
The fine-tuning process on a new target site proves to be relatively efficient in terms of time and computation resources. 
In our experiments, it takes approximately half an hour and utilizes 10G (out of 24G) of GPU memory on an RTX 3090 GPU. More details are discussed in Section~\ref{sec:cost}.
By disentangling image generation via SIG and style translation via SST in the low-dimensional latent space, it is expected to achieve efficient and generalizable MRI harmonization.

%Through the disentanglement of im-age generation and style translation in latent space, the DLEST can achieve efficient and generalizable MRI harmonization.

\subsubsection{Site-Specific MRI Synthesis (SMS)} 
%Alternatively, in step 2, 
In the proposed SST module, we can substitute the original source MRI and encoder $\bm E$ with a random Gaussian noise $z$ and the mapping network $\bm F$ to perform \emph{site-specific MRI synthesis}, as shown in Fig.~\ref{fig_pipeline}~(c). 
In this case, as the network $\bm F$ has already been trained in the SIG to generate latent codes resembling brain MRIs, no further fine-tuning is required. 
The EBM will then translate these randomly generated latent codes into the target style for image synthesis. 
% It's worth noting this ability to synthesize MRI with a specific site style is helpful for training general %deep-
% learning models for downstream %non-disease-specific 
% tasks, especially when real MRI data is limited. 
% {\color{blue}We can say more about the potential applications of our synthetic MRI with consistent style, while other methods such as GAN fail to do this.}
Unlike many MRI synthesis models that only generate MRI with random anatomical structures and styles, our framework can synthesize MRIs with a consistent intensity distribution that matches a given image domain, this can be effectively used as a data augmentation method to pre-train models that require large number of MRI samples without introducing non-biological variations caused by difference in imaging style.

\begin{figure*}[!t]
\setlength{\abovecaptionskip}{-2pt}
\setlength{\belowcaptionskip}{-2pt}
\setlength{\abovedisplayskip}{-2pt}
\setlength{\belowdisplayskip}{-2pt}
\includegraphics[width=1\textwidth]{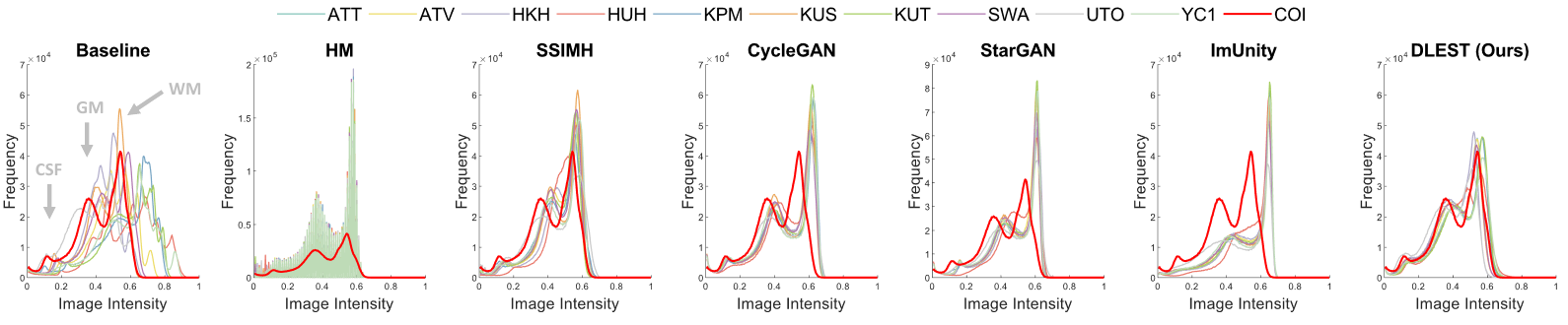}
%% Something like this 
\caption{Histogram comparison of 10 source sites and a target site (COI) across all 9 traveling subjects from the SRPBS dataset. The first plot shows pre-harmonization histograms, while the subsequent plots depict post-harmonization histograms by each competing method and our method DLEST, respectively. WM: white matter; GM: gray matter; CSF: cerebrospinal fluid.} %Three prominent peaks, representing CSF, GM, and WM, are better aligned after harmonization}
\label{fig:histogram}
\end{figure*}

\subsubsection{Implementation} 
%\subsubsection{Collaborative Training.} 
%For fine-tuning both the SST and the SMS modules, we use both the OpenBHB dataset, due to its relatively larger sample size and abundant sites/settings. 
The implementation of SIG incorporates similar settings as outlined in \citep{ALAE}, with structural modifications made to accommodate our specific datasets and tasks. 
Specifically, both the mapping function $\bm{F}$ and the latent discriminator $\bm{D}$ are implemented using multi-layer perceptrons (MLPs). 
The former consists of eight layers, while the latter is composed of three layers. Two networks $\bm{E}$ and $\bm{G}$ in the latent autoencoder are composed of seven encoding/decoding blocks. 
Each of these blocks incorporates an increasing number of filters, ranging from 64 to 256 filters. 
All latent codes have a dimension of 512 in the experiments.
We train the SIG module using an Adam~\citep{kingma2014adam} optimizer %on OPthe entire multi-site dataset 
in an \emph{unsupervised training} scheme. 
Compared to GANs~\citep{CycleGAN,StarGANv2}, SIG’s training cost is lower due to two factors: (1) most of its losses are evaluated in low-dimensional latent space, and (2) no additional cost for cycle-consistency constraints in image space. 
%After completing the training, 
Following the SIG training, we can directly use the latent autoencoder 
%$\bm E$ and $\bm G$ can be directly employed 
in SST training, during which $\bm E$ and $\bm G$  are not updated.

The training of the SST is \emph{efficient in both training time and computational cost} due to four factors.  
(1) The autoencoder $\bm E$ and $\bm G$ do not require additional training, even when deployed on previously unseen data. 
(2) The style translation process, facilitated by EBM, operates in a low-dimensional latent space with a dimension of 512, which is considerably smaller than the image space. 
(3) The energy function $\mathcal{E}_{\theta}$ in EBM is designed to be light-weight, also implemented as an MLP with 2 layers,  %activated by LeakyReLU, 
each having a dimension of 2,048.
(4) The SST can be fine-tuned with a small sample size at low computational cost, which will be discussed in Sections~\ref{sec:target} and ~\ref{sec:cost}. 
In this way, our DLEST can achieve efficient imaging style translation and generalizable MRI generation %and efficient style translation 
through the disentanglement scheme.

%%%%%% -- New Section -- %%%%%%
\section{Experiments}
% more detail about competing method
\subsection{Experimental Settings}
\subsubsection{Competing Methods}\label{sec_competing_methods}
We first compare DLEST with two non-learning methods, including (1) histogram matching (\textbf{HM})~\citep{HM_2014} with multiple experiments with randomly selected reference MRI from the target dataset, and (2) Spectrum Swapping-based Image-level Harmonization (\textbf{SSIMH}), which replaces certain low-frequency region of the source MRI with the corresponding low-frequency spectrum of the reference MRI, with a default swapping threshold 3. We use the DomainATM toolbox~\citep{DomainATM} to carry out two non-learning methods. 
We also compare DLEST against the following state-of-the-art (SOTA) methods for unpaired image-level harmonization. % GAN-based methods. %, with details introduced below 
%{\color{blue}[We can also point out that the competing methods need paired MRIs for training, while our model can be trained using unpaired MRIs? More details on these competing methods can be included here.]}

(1) \textbf{CycleGAN}~\citep{CycleGAN} is initially introduced for natural image-to-image translation. It enhances the capabilities of traditional GANs by integrating an important cycle-consistency constraint. This additional constraint enables the unpaired mapping from a source domain to a target domain. The CycleGAN has been utilized in several MRI harmonization studies \citep{Modanwal_2020_cyclegan,chang_2022_cyclegan} and has become the foundational structure of many subsequent image-level harmonization methods.

(2) \textbf{StarGAN}~\citep{StarGANv2} is initially introduced for multi-domain natural image synthesis task. 
Then, it was employed for MRI harmonization~\citep{StyleGAN}. 
Building upon CycleGAN, the StarGAN method learns a separate latent style code for each target image and injects this code into the source image via adaptive instance normalization.

(3) \textbf{ImUnity}~\citep{ImUnity_2023} is designed specifically for MRI harmonization. It employs a  self-supervised variational autoencoder %variational auto-encoder (VAE) and GAN 
model, coupled with a confusion module for site/scanner-bias unlearning. 
In addition, it includes an optional biological preservation module to preserve clinical information, such as disease labels or diagnostic scores. 
% However, 
In this work, we use the default form of ImUnity without this optional module, since the two datasets only consist of healthy control subjects.

In all three deep learning methods mentioned above, we typically adhere to the default setting of each competing method and make a dedicated effort to ensure other training hyperparameters align closely with the proposed DLEST framework.
In DLEST, the two hyperparameters $\alpha$ and $\beta$ in Eq.~\ref{eq:10} are empirically set as 1 and 100, respectively.
For a fair comparison, 
%In all the three deep learning methods mentioned above, 
we standardize the input size to be $256\times256$ and set the latent dimension to 512, mirroring the specifications of the proposed DLEST. 
%We typically adhere to the default setting of each competing method and make a dedicated effort to ensure other training hyperparameters align closely with ours. 
To train deep learning models (\ie, CycleGAN, StarGAN, ImUnity, and our DLEST), we use the OpenBHB dataset, due to its relatively larger sample size and abundant sites/settings. 
Following the data split in~\citep{OpenBHB}, $32,270$ training slices and $7,570$ validation slices are used, with manual inspection and quality control. % for model training. 

% {\color{red}ComBAT + NeroComBAT}

\subsubsection{Evaluation Tasks}% \& Data Partition}
Four tasks are performed, including 
(1) histogram comparison and feature visualization, 
(2) site classification,  
(3) brain tissue segmentation, and 
(4) site-specific MRI synthesis.  
%For training the SIG module, we use the OpenBHB dataset, due to its relatively larger sample size and abundant sites/settings.  
%For fine-tuning both the SST and the SMS modules, we use both the OpenBHB dataset, due to its relatively larger sample size and abundant sites/settings. 
The SRPBS is used in Tasks 1, 3, and 4 because traveling subjects are available. 
%{\color{red}%Specifically, MR scans of 9 subjects from 11 acquisition sites/settings in SRPBS are used, while one duplicated site (YC2) is removed. 
The OpenBHB is used in Task 2 and Task 4. %, due to its relatively larger sample size and abundant sites/settings. 
%After manual inspection and quality control, 3,227 training scans and 757 validation scans from 58 sites/settings are used.
%}

\subsection{Experimental Results}
 \subsubsection{Task 1: Histogram Comparison \& Feature Visualization}
%\noindent\textbf{Experiment 2: Harmonization Effect on Traveling Subjects} 
% For qualitative evaluation, % the harmonization effect, 
% we compare histograms and MRIs before and after harmonization via DLEST on SRPBS with traveling subjects that have ground truth scans. %is available to assess the ability of our DLEST to perform style translation to a target site. 
% We select the site COI as target since it has relatively low within-site variation, and harmonize the rest 10 sites/settings to COI.
% It can be seen from Fig.~\ref{fig:histogram} that there is increased overlapping between each source site and the target after harmonization. 
% In most cases, three prominent histogram peaks (corresponding to CSF, GM, and WM) are better aligned to those of the target, indicating the effectiveness of our DLEST. % in MRI harmonization. 
% Also, Fig.~\ref{fig_mriShow} suggests that the overall style of harmonized MRIs generated by DLEST is more consistent with target images, compared with competing methods. 
We first qualitatively evaluate our DLEST  by comparing histograms and extracted features of MRIs before harmonization (Baseline) and post-harmonization produced by each competing method in Fig.~\ref{fig:histogram}.  
The validation is performed on SRPBS since each subject has ground-truth scans on multiple acquisition sites.
To select a target site, we compare intra-site variations across each site in the SRPBS dataset by computing the mean peak signal-to-noise ratio (PSNR) for all image pairs within a given site.
Given that the SRPBS encompasses all traveling subjects, each site shares an identical subject cohort, thereby containing consistent anatomical information. 
Consequently, a site yielding a higher intra-site PSNR score suggests a greater similarity in image style within that particular site. 
We treat COI as the target domain due to its low intra-site variation and harmonize the remaining 10 sites/settings to the COI.
The influence of target site selection is elaborated in Section~\ref{sec:target}.

As observed in Fig.~\ref{fig:histogram}, before MRI harmonization (baseline), each site initially exhibits unique intensity distributions. 
In particular, three prominent intensity peaks, WM, GM, and CSF, are not aligned with the target domain. 
%{\color{red}
After harmonization, our DLEST framework demonstrates superior performance in aligning the histogram of the ten source sites with the target COI, particularly excelling in aligning the three main peaks. 
However, the three SOTA deep learning methods (\ie, CycleGAN, StarGAN, and ImUnity) cannot well align source peaks with the target ones, despite showing good peak alignment across the 10 source sites.  
Our success may be attributed to the model's ability to learn the latent data distribution of the entire target domain via EBM, while the three deep learning methods do not effectively account for intra-site variation within the target domain. %for harmonization. 
On the other hand, the two non-learning methods cannot produce good performance. 
For example, the HM method, despite achieving %satisfactory
reasonable peak alignment with the target, exhibits a non-smooth artifact in the post-harmonization histogram. 
This could be due to the significant intensity disparity between the source MRIs and the target/reference MRI~\citep{li_2021_impact}, caused by potential site effects. 
\if false
Even though the three deep learning methods (\ie, CycleGAN, StarGAN, and ImUnity) show good peak alignment across the 10 source sites 
Other methods, including deep learning methods like CycleGAN, StarGAN, and ImUnity, show good peak alignment across the 10 source sites but fail to align with the target peaks.
\fi 
%The post-harmonization histogram, generated using the HM method, displays a non-smooth artifact despite satisfactory peak alignment with the target. 
%This lack of smoothness could be attributed to the significant intensity disparity between the source MRIs and the target reference MRI \citep{li_2021_impact}. 
% As all sites share the same subject cohort, this disparity is likely due to site effect, such as variations in scanner field strength. 
% The HM method, which maps the source histogram to the reference MRI’s histogram bins, does not take spatial information into account, potentially contributing to this non-smoothness. 
% While all other methods demonstrate good peak alignment across the 10 source sites, deep learning methods like CycleGAN, StarGAN, and ImUnity outperform the non-learning-based SSIMH method. 
% However, only our DLEST framework closely aligns the histogram of source sites with the target, particularly excelling in aligning the three main peaks. 
% The superiority of our result lies in learning the latent data distribution of the entire target site, rather than relying on one single reference MRI.
%}

For feature visualization, we use a pre-trained ResNet18~\citep{ResNet} model to extract features of the original MRIs and harmonized ones (via DLEST) 
across all sites and subjects. 
We then apply principal component analysis (with 100 principal components) %n\_component=20) 
to reduce the dimensionality of the extracted image features, and then utilize the t-SNE~\citep{van2008visualizing} to reduce the dimensionality of these features further and project them into a 2D space. 
Finally, we color-code each site with a unique color.
The visualization results in Fig.~\ref{fig:cluster} show that the extracted image features are initially clustered into 11 distinct groups. 
After harmonization by DLEST, these clusters become less distinguishable, implying that our method effectively eliminates site-related variations. % in harmonized MRIs.
%our DLEST framework effectively eliminates site-related variations, leading to less distinguishable site clusters post-harmonization.
%} % which implies......

\begin{figure}[!t]
\setlength{\abovecaptionskip}{-2pt}
\setlength{\belowcaptionskip}{-2pt}
\setlength{\abovedisplayskip}{-2pt}
\setlength{\belowdisplayskip}{-2pt}
\includegraphics[width=0.49\textwidth]{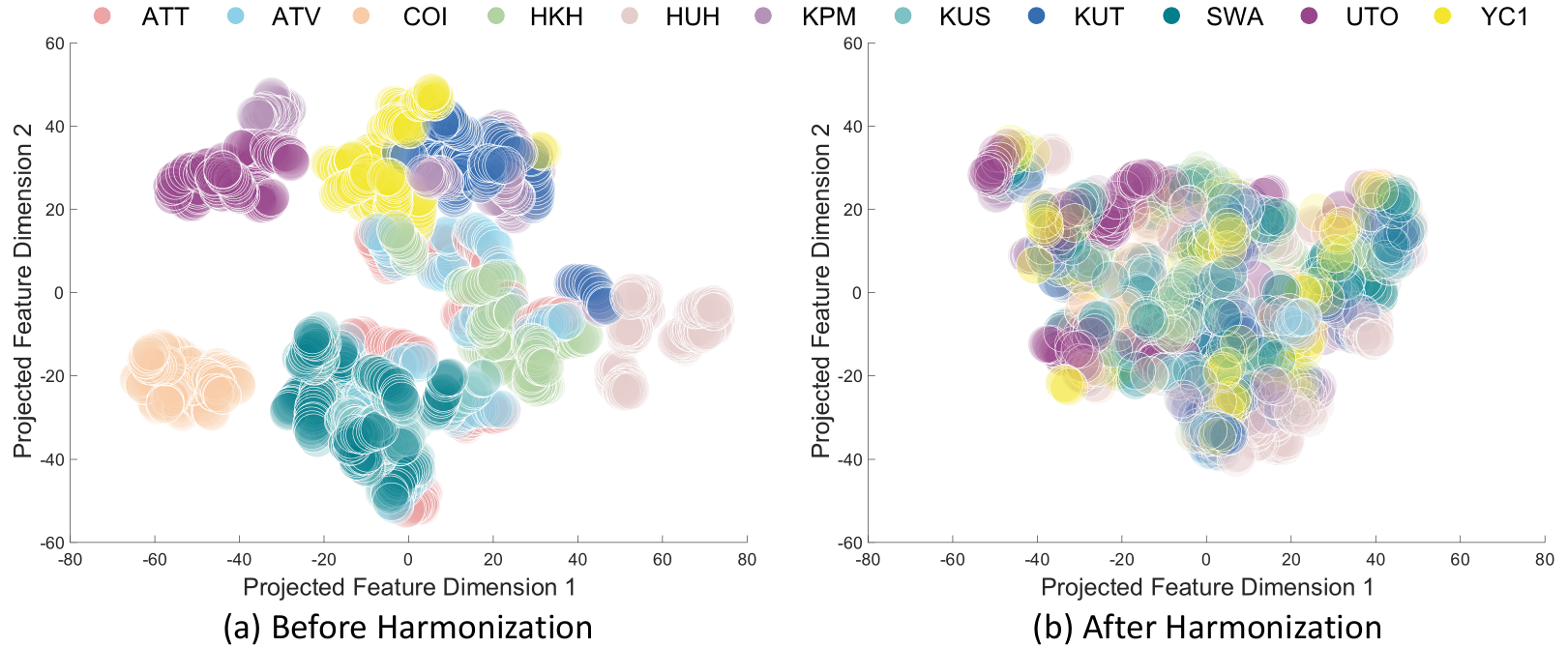}
%% Something like this 
\caption{Visualization of MRI features of 11 sites across 9 subjects from SRPBS, with MRI (a) before and (b) after harmonization by DLEST. Each color denotes a specific site, while each point denotes extracted features from an MRI slice of a specific subject.} 
\label{fig:cluster}
\end{figure}

\subsubsection{Task 2: Site Classification}
%Experiment 1: Harmonization Effect on Site Classification.} 
 In this task, we conduct a quantitative evaluation to determine the effectiveness of a harmonization method in eliminating site-related, non-biological variations. 
 We begin by selecting a site (ID: 17) from  OpenBHB with minimal intra-site variation as our target site. 
 We then harmonize MRIs from the rest 57 source sites to this target domain using each method. 
 Following this, we train a site classifier on OpenBHB training set (with $3,227$ MRIs) and validate it on the validation set (with $757$ MRIs). 
 The site classifier is implemented based on the ResNet18~\citep{ResNet} architecture. 
 Its performance on raw MRI data is recorded as the Baseline. 
 %Our evaluation utilizes 
 Five metrics are used: balanced accuracy (BACC), accuracy (ACC), area under the ROC curve (AUC), F1 score (F1), and sensitivity (SEN).
% In this task, we  
% %This experiment aims to 
% quantitatively assess whether a harmonization method can effectively remove non-biological variations. % associated with each acquisition site/setting. 
% We first harmonize all data from 58 sites in OpenBHB to a pre-selected target site (ID: 17) using a specific method. 
% We then train a ResNet18~\citep{ResNet} as the site classifier on the training set of OpenBHB and validate it on the validation set.  
% We record the performance of ResNet18 on raw data as Baseline. 
% Five metrics are used: balanced accuracy (BACC), 
% accuracy (ACC), area under the ROC curve (AUC), F1 score (F1), and sensitivity (SEN). 
% are used to evaluate the site classifier. 
The results of site classification are reported in Table~\ref{tab:1}, where a lower value signifies a method's better capability to eliminate site-related variations, thereby making it challenging for the site classifier to learn site-related features that could differentiate each site.

As indicated in Table~\ref{tab:1}, our DLEST method consistently surpasses the five competing methods by a substantial margin.
% The site classification results based on data harmonized by five methods are reported in Table~\ref{tab:1}. 
% Here, a lower value indicates a stronger ability of a method to remove site effects, making it more difficult for ResNet18 to learn non-biological features that can distinguish each site. 
% Table~\ref{tab:1} suggests that our DLEST consistently outperforms all competing methods by a large margin. 
For instance, after the data is harmonized by DLEST, there is a decrease in BACC by $0.340$ compared to the Baseline, and a decrease of $0.153$ when compared to the second-best StarGAN method. 
When compared with the three SOTA methods (\ie, CycleGAN, StarGAN, and ImUnity), our method consistently yields superior results across five metrics. 
This result is consistent with the qualitative result in Task 1 (see Fig.~\ref{fig:histogram}). 
% For example, the BACC decreases by  when trained on data harmonized by DLEST compared to raw data, and decreases by $0.153$ compared to the second-best method (\ie, StarGAN). 
% Compared with three SOTA methods (\ie, CycleGAN, StarGAN, and ImUnity), our DLEST yields very good results in terms of five metrics (\ie, BACC, ACC, AUC, F1, and SEN). 

Several factors may contribute to the significant performance difference between our DLEST and the three SOTA methods. 
Unlike CycleGAN, which relies solely on adversarial learning between discriminators and the generators in image space to generate images resembling the target, our DLEST achieves style translation by directly manipulating the latent code of an image via EBM. 
This direct manipulation of latent code allows our DLEST to have fine-grained control over style translation. 
Furthermore, DLEST captures the underlying data distribution of the entire target site when translating an image to the target style. 
In contrast, StarGAN treats each target image as a unique ``domain'' with its own style and depends on a single reference image during style translation. 
Similarly, while ImUnity learns an unbiased latent representation devoid of site information, it still necessitates a single reference image for ``contrast correction''. 
These quantitative results further suggest the superiority of DLEST in capturing the latent distribution of the target domain for effective style translation.

% This implies that our method generalizes well to multi-site MRIs with different data distributions. 
% The possible reason could be that our method is able to capture the underlying distribution of source and target sites in latent code space, instead of relying on a single reference/target image. 
%treating each target image as a unique 'domain' with its own style and relying on a single reference image during the style translation 

\if false
The result in Table \ref{tab:1} shows the performance of the site classifier when trained on raw data, as well as on data harmonized by each method. A lower score indicates stronger ability of a method to remove site effects, making it more difficult for the site classifier to learn non-biological features that can distinguish each site. 
Among the competing methods, HM showed a decent harmonization performance, even outperforming two GAN-based methods, possibly because it directly aligns the global intensity distributions of source and target images, without encoding the images into latent representations. 
This makes it less affected by the intra-site content and style variations. 
Nonetheless, our proposed DLEST framework outperformed all competing methods by a significant margin. 
For example, the BACC of the site classifier decreased by $0.448$ when trained on data harmonized by DLEST compared to raw data, and decreased by $0.129$ compared to HM. 
One reason for the significant performance difference between our DLEST and the two SOTA GAN-based methods is that, unlike StarGAN, our model is able to translate an image to the target style by capturing the underlying data distribution of the entire target site, instead of treating each target image as a unique ``domain'' with its own style and relying on a single reference image during the style translation~\citep{StyleGAN}. Compared to the CycleGAN, our DLEST achieves style translation by directly manipulating the latent code of an image via EBM, whereas the CycleGAN relies solely on the adversarial learning between the discriminator and the generator to generate images that are similar to the target style, without directly manipulating its latent code. 
The latent code manipulation allows our DLEST to have fine-grained control over the style translation, as opposed to mapping one domain to another entirely. 
\fi

\begin{figure*}[!t]
\setlength{\abovecaptionskip}{-3pt}
\setlength{\belowcaptionskip}{-2pt}
\setlength{\abovedisplayskip}{-2pt}
\setlength{\belowdisplayskip}{-2pt}
\centering
\includegraphics[width=1\textwidth] {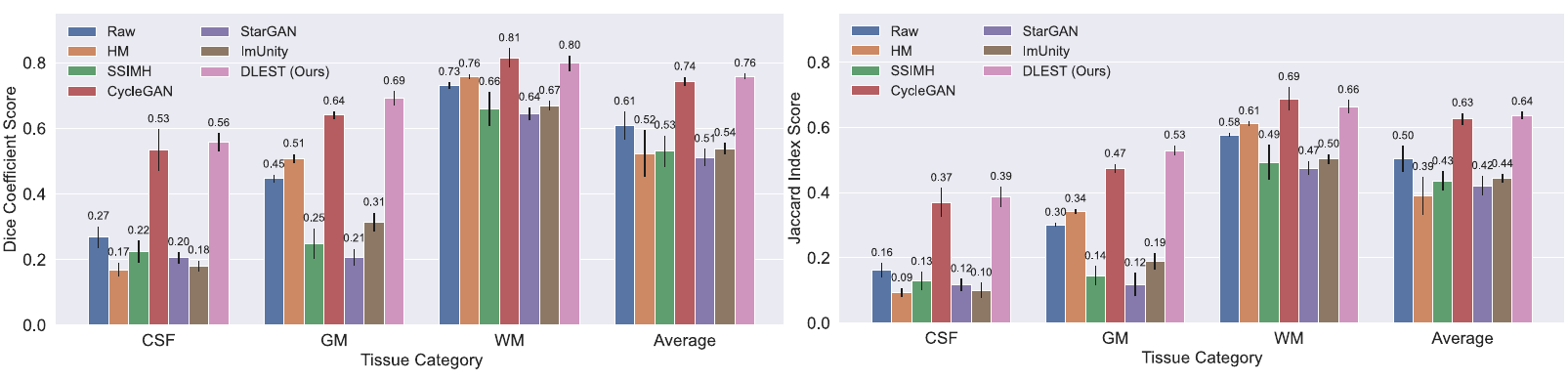}
\caption{Segmentation results of %3 tissues (\ie,
CSF, GM, and WM on SRPBS in terms of 
(left) Dice coefficient and (right) Jaccard index. Each U-Net is trained on the target site COI and validated on the source site HUH from SRPBS.% We can say more in the caption.[(ours) should be (Ours)]}
}
%with of the UNet trained on the target site COI and validated on the source site HUH.}% Validation performance on raw data, as well as data harmonized by each method, are compared.}
\label{fig:unet}
\end{figure*}

\begin{figure}[t]
\setlength{\abovecaptionskip}{0pt}
\setlength{\belowcaptionskip}{-2pt}
\setlength{\abovedisplayskip}{-2pt}
\setlength{\belowdisplayskip}{-2pt}
\centering
\includegraphics[scale=0.90] {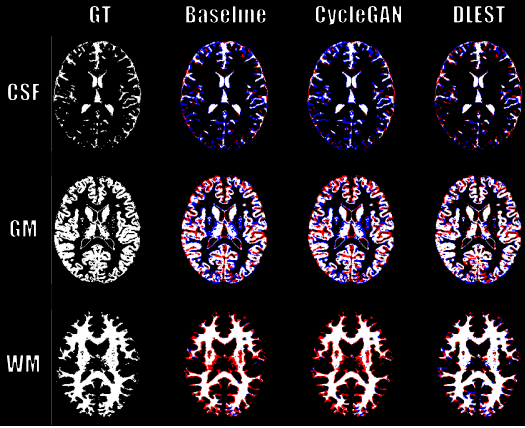}
\caption{Segmentation of brain tissues (\ie, CSF, GM, WM) using unharmonized MRI data (Baseline) and data harmonized by CycleGAN (top competing method) and DLEST (ours). White pixels denote the correct segmentation, red pixels denote under-segmentation, and blue pixels denote over-segmentation.}
\label{fig:sample_seg}
\end{figure}

\begin{table}[!t]
\setlength{\abovecaptionskip}{0pt}
\setlength{\belowcaptionskip}{-0pt}
\setlength{\abovedisplayskip}{-0pt}
\setlength{\belowdisplayskip}{-0pt}
\centering
\caption{%Harmonization effects of all methods on site classification performance.
Multi-class (58 categories) site classification results on OpenBHB, 
where `*' denotes that the difference between a specific competing method and DLEST is statistically significant ($p<0.05$ via paired $t$-test).}
%Site classification results produced by six different methods on the OpenBHB dataset (mean$\pm$standard deviation), where `*' denote the BACC difference between a specific competing method and the DLEST is statistically significant ($p\leq0.05$ via paired $t$-test).}%  with lower values indicating better results.} %removal of site-related variations.}
 % \scriptsize
% \footnotesize
\tiny
\setlength{\tabcolsep}{5pt} % to reduce space between columns
\begin{tabular}{l|ccccc}
\toprule
%\multirow{2}{*}{Method} &\multicolumn{5}{c}{Metrics}\\ \cmidrule(l){2-6}
% \cmidrule(l){6-9}
% \cmidrule(l){10-13}
Method        & {BACC$\downarrow$} & {ACC$\downarrow$}    & {AUC$\downarrow$}  & {F1$\downarrow$}  & {SEN$\downarrow$}\\
\midrule
Baseline$^*$   & $0.676$$\pm$$0.137$ & $0.722$$\pm$$0.147$   & $0.979$$\pm$$0.022$  & $0.713$$\pm$$0.146$  & $0.722$$\pm$$0.147$ \\
HM$^*$      & $0.619$$\pm$$0.133$ & $0.696$$\pm$$0.137$     & $0.981$$\pm$$0.019$  & $0.685$$\pm$$0.129$  & $0.696$$\pm$$0.137$ \\
SSIMH$^*$   & $0.673$$\pm$$0.083$ & $0.659$$\pm$$0.076$     & $0.983$$\pm$$0.008$  & $0.656$$\pm$$0.079$  & $0.659$$\pm$$0.076$ \\
CycleGAN$^*$  & $0.588$$\pm$$0.081$ & $0.660$$\pm$$0.088$   & $0.974$$\pm$$0.007$  & $0.645$$\pm$$0.088$  & $0.660$$\pm$$0.088$ \\
StarGAN$^*$   & $0.489$$\pm$$0.201$ & $0.562$$\pm$$0.223 $  & $0.951$$\pm$$0.047$  & $0.543$$\pm$$0.215$  & $0.562$$\pm$$0.223$ \\
ImUnity$^*$ & $0.569$$\pm$$0.074$ & $0.658$$\pm$$0.066$ & $0.978$$\pm$$0.008$ & $0.636$$\pm$$0.066$ & $0.658$$\pm$$0.066$ \\
\textbf{DLEST~(Ours)}   & $\bm{0.336}$$\pm$$0.057$ & $\bm{0.483}$$\pm$$0.065$ & $\bm{0.938}$$\pm$$0.020$ & $\bm{0.474}$$\pm$$0.069$  & $\bm{0.483}$$\pm$$0.065$ \\
\bottomrule
\end{tabular}
\label{tab:1}
\end{table}

\if false
\begin{figure}[t]
\setlength{\abovecaptionskip}{0pt}
\setlength{\belowcaptionskip}{-2pt}
\setlength{\abovedisplayskip}{-2pt}
\setlength{\belowdisplayskip}{-2pt}
\centering
\includegraphics[scale=0.95] {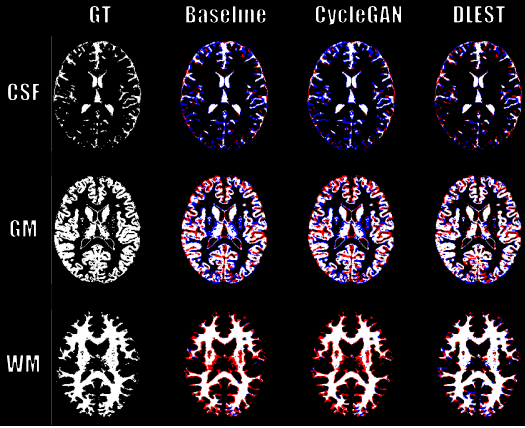}
\caption{Sample U-Net segmentation of brain tissues (CSF, GM, WM) using unharmonized data (Baseline) and data harmonized by CycleGAN (top competing method) and DLEST (ours). White pixels denote the correct segmentation, red pixels denote under-segmentation, and blue pixels denote over-segmentation.}
\label{fig:sample_seg}
\end{figure}
\fi

\begin{figure*}[t]
\setlength{\abovecaptionskip}{1pt}
\setlength{\belowcaptionskip}{-2pt}
\setlength{\abovedisplayskip}{-2pt}
\setlength{\belowdisplayskip}{-2pt}
\centering
\includegraphics[width=1\textwidth]{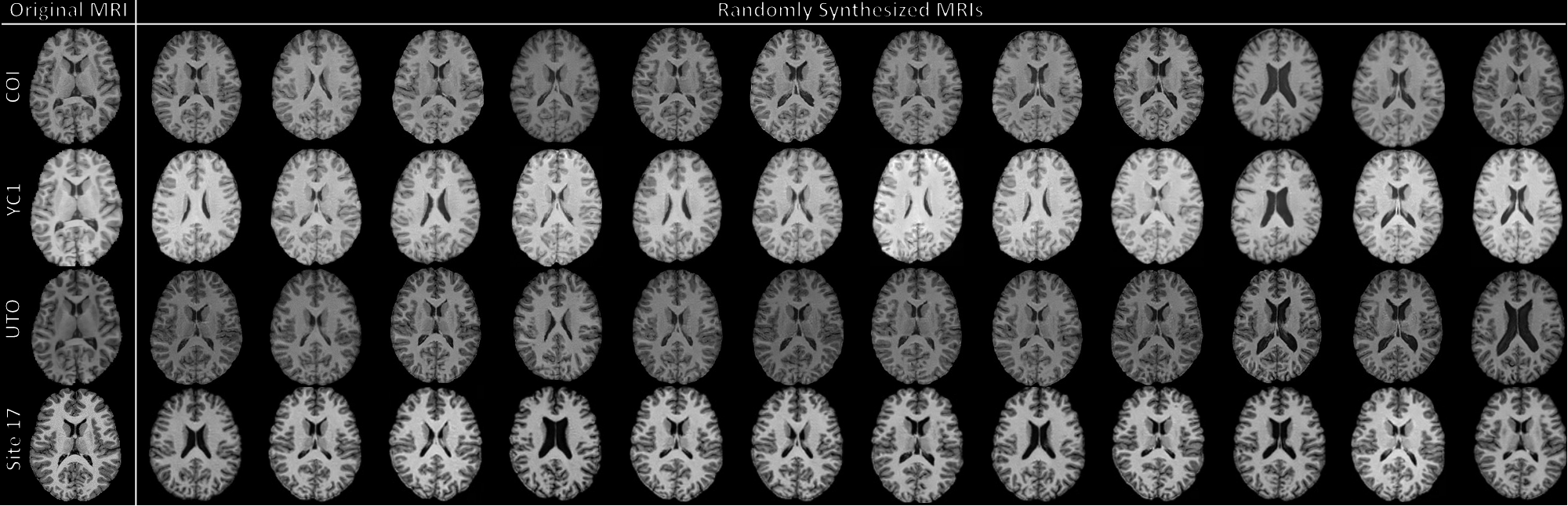}
\caption{Synthesized MRI with site-specific styles of 3 sites from SRPBS (top 3 rows) and 1 site from OpenBHB (bottom row). 
% {\color{blue}We can say more in the caption.}
The synthesized MRIs have diverse anatomical structures with consistent image styles that match the reference MRI from a specific site.}
\label{fig_synShow}
\end{figure*}

\subsubsection{Task 3: Brain Tissue Segmentation}

This experiment is designed to assess the effectiveness of the DLEST on a downstream brain tissue (\ie, WM, GM, and CSF) segmentation task. We begin by selecting a source site (\ie, HUH) and a target site (\ie, COI) within the SRPBS dataset and harmonize the source MRIs to the target domain using DLEST and each competing method.
Similar to~\citep{StyleGAN}, a U-Net~\citep{UNet} segmentation model is trained on MRIs from the target site (COI) and directly validated on the source site (HUH). 
As our primary objective is to assess the impact of each harmonization method on tissue segmentation, the quality of the ground-truth segmentation is not our main concern. 
We thus utilize the segmentation masks obtained using the FAST pipeline in FSL as our ground truth for UNet training, providing a fair basis for comparing each method.
% This experiment aims to further validate the impact of DLEST on a downstream image analysis task (\ie, brain tissue segmentation). We first harmonize a selected source site (\ie, HUH) from SRPBS to a target site space (\ie, COI) using each aforementioned harmonization method. We then train a U-Net~\citep{UNet} segmentation model on images from COI and directly validate the trained U-Net on HUH. The auto-segmentation outputs from FSL~\citep{FSL} are used as the ground truth. The U-Net performance on unharmonized HUH data serves as the baseline. Dice coefficient and Jaccard index are used to assess U-Net performance. 
% The class-wise and average brain tissue segmentation results are shown in Fig.~\ref{fig:unet}.
For each harmonization method, the U-Net is trained for 100 epochs, with the mean and standard deviation of the validation Dice coefficient and the Jaccard index reported for the final 40 epochs.

\if false
{\color{red}Figure~\ref{fig:sample_seg} presents a comparative visualization of the brain tissue segmentation results. The U-Nets were trained on both unharmonized data (\textbf{Baseline}) and data harmonized by the top competitor \textbf{CycleGAN} and our \textbf{DLEST} framework. 
In this figure, the segmentation map uses white pixels to indicate accurate segmentation, while red and blue pixels represent under-segmentation and over-segmentation, respectively. 
Notably, the U-Net trained on DLEST-harmonized data demonstrated significant improvements, particularly in reducing under-segmentation of white matter near tissue boundaries and over-segmentation of gray matter near the thalamus region.}
\fi

% Quantitatively, as shown in Fig.~\ref{fig:unet}, 
% although the source and the target site share the same subject cohort, directly applying the UNet that is trained on site COI to raw data in site HUH yields suboptimal segmentation results due to the site effect.
Figure \ref{fig:unet} quantitatively illustrates that despite the source and target sites sharing the same subject cohort, the direct application of a UNet model trained on the site COI to raw data from site HUH results in less than optimal segmentation. This suggests that site-related variations hinder the generalizability of the segmentation model.
Our DLEST method consistently surpasses all non-learning methods (\ie, HM and SSIMH), suggesting that the latent space style translation is more effective than merely aligning the global intensity distribution. 
%This result corroborates the histogram alignments shown in Fig.~\ref{fig:histogram}. 
%In comparison to the three SOTA deep learning methods, 
Compared with the three SOTA methods, our DLEST consistently delivers improved segmentation performance in terms of average segmentation results specifically in the segmentation of CSF and GM. 
While CycleGAN slightly outperforms DLEST in WM segmentation, % by a slight margin of 0.017, 
our DLEST exhibits a significantly lower standard deviation. 
This indicates that the DLEST produces a more stable performance compared with CycleGAN.

In addition, Fig.~\ref{fig:sample_seg} presents the comparative visualization of brain tissue segmentation results. The first column displays the ground truth (\textbf{GT}) segmentation maps, derived from FSL. 
The subsequent columns represent the results of three identically structured U-Nets, all trained on the target domain (\ie,~COI) MRIs. 
In particular, the \textbf{Baseline} column shows validation results on unharmonized source domain (\ie,~HUH) MRIs. 
The \textbf{CycleGAN} column presents the validation results on the source domain MRIs, post-harmonization to the target domain using the top competing method (\ie, CycleGAN). 
Similarly, the \textbf{DLEST} column depicts the validation results on source domain MRIs harmonized using our method.
In this figure, the segmentation map uses white pixels to indicate accurate segmentation, while red and blue pixels represent under-segmentation and over-segmentation, respectively. 
Notably, the U-Net trained on DLEST-harmonized data demonstrated significant improvements, particularly in reducing under-segmentation of white matter near tissue boundaries and over-segmentation of gray matter near the thalamus region.

\subsubsection{Task 4: MRI Synthesis}
This task aims to demonstrate our framework's ability to synthesize %realistic 
brain MRIs with site-specific styles. 
We select three sites from SRPBS and one site from OpenBHB for illustration. 
We evaluate %the quality of 
the synthesized MRIs qualitatively via sample visualization
and quantitatively using structural similarity index measurement (SSIM)~\citep{SSIM}, peak signal-to-noise ratio (PSNR), root mean square error (RMSE), and intensity Pearson correlation coefficient (PCC). 
In our analysis, we apply a binary brain mask to eliminate any influence of the background pixels when calculating each metric.
As shown in Fig.~\ref{fig_synShow}, our DLEST is able to synthesize new brain MRIs with diverse brain shapes and anatomical structures while preserving the site-specific styles that match the given reference MRI scans. 
This latent space MRI synthesis is cost-efficient and helpful
% in training general learning models when the real MRI samples are limited.
in training non-disease-specific learning models that require large quantities of MRI samples while the actual MRI data is limited, such as brain segmentation models.

\if false
For quantitative evaluation, we report the mean and the standard deviation of image-level metrics for synthesized MRIs at each site in Table~\ref{tab:syn_quant}. 
%Table~\ref{tab:syn_quant} presents the mean and the standard deviation of image-level metrics for the synthesized MRI at each site. 
Metrics under the \textbf{Original} category are derived from comparisons between each original image and every other original image within the same site. 
The \textbf{Synthesized} category metrics are obtained by comparing each synthesized image with all original images from the same site. The marginal differences observed across all metrics between Original and Synthesized image pairs suggest that our DLEST framework can effectively generate synthetic MRIs that maintain consistent styles and quality comparable to the original images.
\fi 

% \usepackage{multirow}
% \usepackage{graphicx}
\begin{table}[!t]
\setlength{\abovecaptionskip}{0pt}
\setlength{\belowcaptionskip}{-0pt}
\setlength{\abovedisplayskip}{-2pt}
\setlength{\belowdisplayskip}{-2pt}
% \scriptsize
\caption{Quantitative result of the image-level metrics. The term ``Original MRI'' denotes mean and standard deviation between every pair of original images; and ``Synthesized MRI''  denotes mean and standard deviation between every pair of original and synthesized images.}
\label{tab:syn_quant}
\tiny
\resizebox{\columnwidth}{!}{%
\begin{tabular}{l|l|cccc}
\toprule
  Site ID                                       &           Data                       & SSIM$\uparrow$        & PSNR$\uparrow$          & RMSE$\downarrow$         &PCC$\uparrow$\\ 
\midrule
\multicolumn{1}{l|}{\multirow{2}{*}{COI}} & \multicolumn{1}{l|}{Original MRI}    & $0.900$$\pm$$0.025$& $27.206$$\pm$$2.266$& $0.045$$\pm$$0.010$&$0.916$$\pm$$0.037$\\
\multicolumn{1}{l|}{}                     & \multicolumn{1}{l|}{Synthesized MRI} & $0.901$$\pm$$0.019$& $26.563$$\pm$$1.817$& $0.048$$\pm$$0.010$&$0.925$$\pm$$0.028$\\ 
\midrule

\multicolumn{1}{l|}{\multirow{2}{*}{YC1}} & \multicolumn{1}{l|}{Original MRI}    & $0.900$$\pm$$0.029$& $25.386$$\pm$$2.672$& $0.056$$\pm$$0.015$&$0.933$$\pm$$0.033$\\
\multicolumn{1}{l|}{}                     & \multicolumn{1}{l|}{Synthesized MRI} & $0.903$$\pm$$0.019$& $25.584$$\pm$$1.852$& $0.054$$\pm$$0.011$&$0.942$$\pm$$0.023$\\ 
\midrule

\multicolumn{1}{l|}{\multirow{2}{*}{UTO}} & \multicolumn{1}{l|}{Original MRI}    & $0.904$$\pm$$0.026$& $27.041$$\pm$$2.572$& $0.046$$\pm$$0.011$&$0.905$$\pm$$0.041$\\
\multicolumn{1}{l|}{}                     & \multicolumn{1}{l|}{Synthesized MRI} & $0.905$$\pm$$0.018$& $27.249$$\pm$$1.680$& $0.044$$\pm$$0.009$&$0.914$$\pm$$0.029$\\ 

\midrule
\multicolumn{1}{l|}{\multirow{2}{*}{Site 17}} & \multicolumn{1}{l|}{Original MRI} & $0.879$$\pm$$0.022$& $25.176$$\pm$$1.892$& $0.056$$\pm$$0.011$&$0.899$$\pm$$0.036$\\
\multicolumn{1}{l|}{}                     & \multicolumn{1}{l|}{Synthesized MRI} & $0.873$$\pm$$0.015$& $24.918$$\pm$$1.270$& $0.057$$\pm$$0.009$&$0.898$$\pm$$0.031$\\ 
\bottomrule
\end{tabular}%
}
\end{table}

For quantitative evaluation, we report the mean and the standard deviation of image-level metrics for synthesized MRIs at each site in Table~\ref{tab:syn_quant}. 
%Table~\ref{tab:syn_quant} presents the mean and the standard deviation of image-level metrics for the synthesized MRI at each site. 
Results under the \textbf{Original} category are derived from comparisons between each original image and every other original image within the same site. 
The \textbf{Synthesized} category results are obtained by comparing each synthesized image with all original images from the same site. 
% The marginal differences observed across all metrics between Original and Synthesized image pairs suggest that our DLEST framework can effectively generate synthetic MRIs that maintain consistent styles and quality comparable to the original images.
The negligible disparity across all four metrics between the original-original and original-synthesized image pairs implies that the synthesized images uphold the same quality as their original counterparts. Moreover, the slightly increased PCC observed in most sites indicates that the synthesized MRI slices exhibit a consistent style and fewer variations when compared to the original MRIs from the respective sites.

\if false
\begin{figure*}[t]
\setlength{\abovecaptionskip}{1pt}
\setlength{\belowcaptionskip}{-2pt}
\setlength{\abovedisplayskip}{-2pt}
\setlength{\belowdisplayskip}{-2pt}
\centering
\includegraphics[width=0.99\textwidth]{TMI_fig_synthetic_v3.pdf}
\caption{Synthesized MRI with site-specific styles of 3 sites from SRPBS (top 3 rows) and 1 site from OpenBHB (bottom row). }
\label{fig_synShow}
\end{figure*}
\fi 

\begin{figure}[t]
\setlength{\abovecaptionskip}{0pt}
\setlength{\belowcaptionskip}{-2pt}
\setlength{\abovedisplayskip}{-2pt}
\setlength{\belowdisplayskip}{-2pt}
\centering
\includegraphics[width=0.5\textwidth] {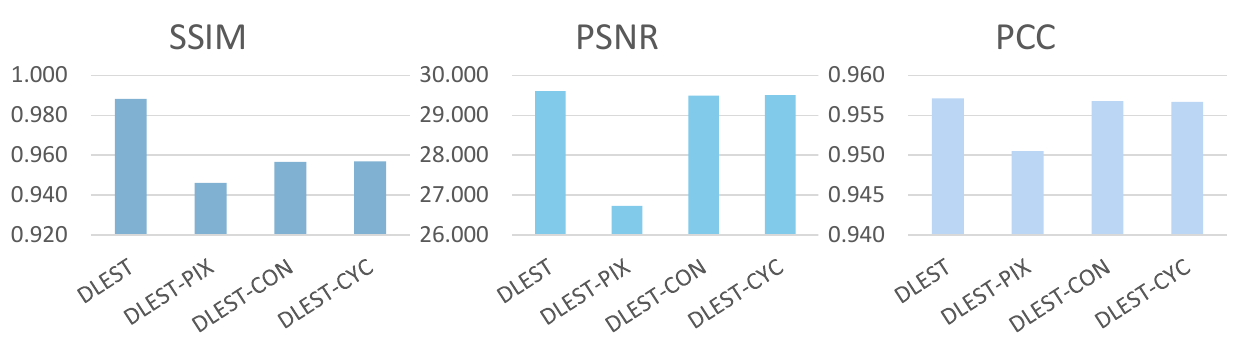}
\caption{MRI quality assessment results achieved by DLEST and its variants on traveling subjects from SRPBS.}
\label{fig_ablation}
\end{figure}

%%%%%% -- New Section -- %%%%%%
\vspace{-8pt}
\section{Discussion}
\subsection{Ablation Study}
We investigate three downgraded variants of the proposed DLEST framework, namely 
(1) \textbf{DLEST-PIX}, which excludes pixel-wise loss in Eq.~\eqref{eq:5} during the SIG module training, 
(2) \textbf{DLEST-CON}, which omits the latent content loss in Eq.~\eqref{eq:10}, and 
(3) \textbf{DLEST-CYC}, which does not incorporate the latent cycle-consistency loss in Eq.~\eqref{eq:10} when training the SST module for style translation. 
We train DLEST and its three variants on the SRPBS traveling subject dataset and evaluate their harmonization results using three image quality metrics: SSIM, PSNR, and PCC.
As indicated in Fig.~\ref{fig_ablation}, 
while all ablations negatively impact the result, 
DLEST-PIX particularly yields the worst results, compared with its two counterparts. 
This implies that the pixel-wise, latent content and cycle-consistency losses are necessary to ensure a satisfying harmonization outcome, and especially, the pixel-wise loss plays a crucial role in preserving image content in DLEST. 
%This implies that pixel-wise loss plays a crucial role in preserving image content in the proposed DLEST and the latent content loss and latent cycle-consistency loss are also necessary to ensure a satisfying harmonization outcome.

\begin{figure*}[t]
\setlength{\abovecaptionskip}{0pt}
\setlength{\belowcaptionskip}{-2pt}
\setlength{\abovedisplayskip}{-2pt}
\setlength{\belowdisplayskip}{-2pt}
\centering
\includegraphics[width=1\textwidth] {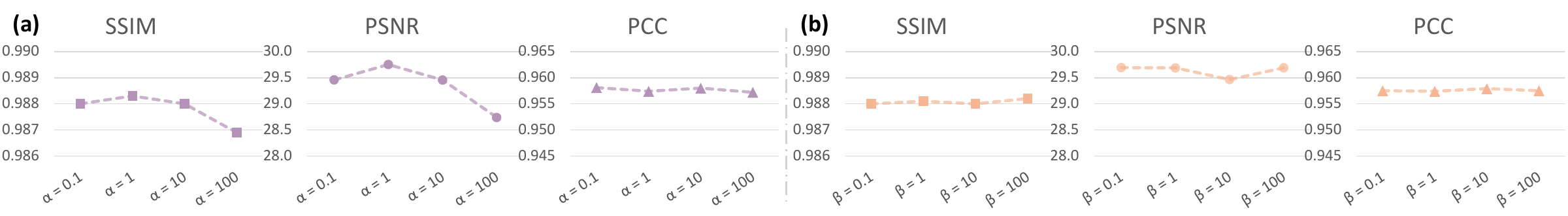}
\caption{Results of DLEST with different values of (a) $\alpha$ and (b) $\beta$ on the SRPBS traveling subject dataset.}
\label{fig_parameter}
\end{figure*}

\subsection{Influence of Hyperparameters}
We conduct several rounds of hyperparameter tuning to examine the influence of hyperparameters $\alpha$ and $\beta$ in Eq.~\eqref{eq:10} on the harmonization effect. This involves varying one parameter while maintaining the other constant. 
The results in Fig.~\ref{fig_parameter} suggest that the model's performance improves when the latent content loss
has a balanced weight (\ie, $\alpha=1$). 
% in the overall loss function. 
Additionally, the performance of DLEST appears to be less sensitive to variations in $\beta$ with optimal performance at $\beta=100$.

% To investigate the impact of hyperparameters $\alpha$ and $\beta$ in Eq.~\eqref{eq:10} on the harmonization effect, 
% we performed several rounds of hyperparameter tuning, varying one parameter while keeping the other unchanged. 
% We use two image-level metrics to assess the harmonization: structural similarity index measurement (SSIM)~\citep{SSIM} and peak signal-to-noise ratio (PSNR). 
% The result in Fig.~\ref{fig_parameter} (b-c) indicates that the model would benefit when the latent content loss carries more weight (\ie, with larger $\alpha$) in the overall loss function. 
% Besides, the performance of DLEST is less affected by the choice of $\beta$.

\if false
\begin{figure*}[t]
\includegraphics[width=0.98\textwidth]{Histogram_v8_1row.pdf}
%% Something like this 
\caption{Histogram comparison of 10 source sites and a target site (COI) from the SRPBS dataset. The first plot shows pre-harmonization histograms, while the subsequent plots depict post-harmonization histograms by each competing method and our method DLEST, respectively.} %Three prominent peaks, representing CSF, GM, and WM, are better aligned after harmonization}
\label{fig:histogram}
\end{figure*}

\begin{figure}[t]
\includegraphics[width=0.49\textwidth]{Clustering_v5.pdf}
%% Something like this 
\caption{Clustering of all 11 sites across all 9 subjects (left) before and (right) after harmonization by DLEST on SRPBS.} 
\label{fig:cluster}
\end{figure}
\fi

\if false
\begin{figure*}[t]
\includegraphics[width=0.98\textwidth]{fig_imageShow_v2.pdf}
\caption{MRIs of 3 subjects in SRPBS dataset, harmonized from HUH to COI by five methods.}
\label{fig_mriShow}
\end{figure*}
\fi 

\begin{figure*}[t]
\setlength{\abovecaptionskip}{0pt}
\setlength{\belowcaptionskip}{-2pt}
\setlength{\abovedisplayskip}{-2pt}
\setlength{\belowdisplayskip}{-2pt}
\centering
\includegraphics[width=1\textwidth] {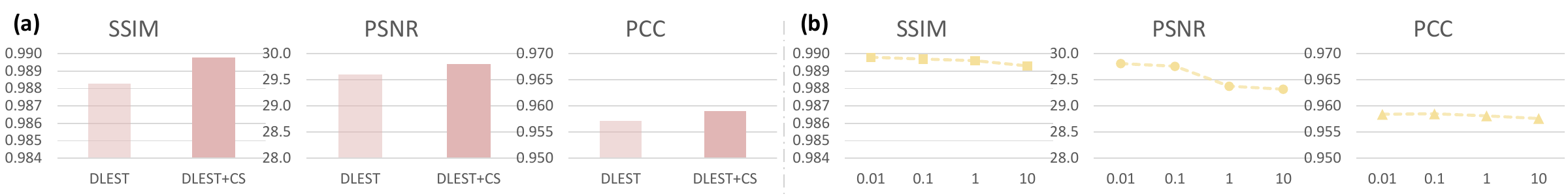}
\caption{(a) Comparison of DLEST and DLEST+CS, and (b) results of DLEST+CS with different style/content ratios on the SRPBS traveling subject dataset.}
\label{fig_csloss}
\end{figure*}

\subsection{Influence of Content Loss}
\label{sec:csloss}
%Our initial implementation 
The proposed DLEST uses a latent content loss in Eq.~\eqref{eq:8} %, as suggested in \citep{zhao2021unpaired}, 
to constrain the change of latent information while minimizing the $\mathcal{L}_{EBM}$ during style translation. 
This implicit regularization in latent space may cause slight alterations in anatomical details. % in some rare cases during our experiments. 
To this end, we compare our method with its variant \textbf{DLEST+CS} that utilizes a separate style and content loss~\citep{gatys2016image} to \emph{explicitly enforce content preservation} while allowing the style to be translated freely. 
In DLEST+CS, a pre-trained VGG19~\citep{simonyan2014VGG} is used to extract feature maps of given MRI slices. 
The \emph{content loss} is calculated as the $l_2$ distance between the extracted feature maps of original MRI and the translated MRI in specific layers. 
The \emph{style loss} is computed as the weighted average of  $l_2$ distance between Gram matrices~\citep{gatys2016image} of the target and translated MRIs in each layer. 
The Gram matrix is a correlation matrix between each feature map, where each entry $\mathcal{G}_{ij}$ is the inner product of vectorized feature maps $i$ and $j$. 
By including feature correlations of multiple layers, we obtain a multi-scale representation of the input image, which captures its texture characteristics without content information~\citep{gatys2016image}.

As shown in Fig.~\ref{fig_csloss}~(a), compared with DLEST, its variant DLEST+CS achieves slight improvement across all metrics, implying enhanced content preservation and style translation. 
We also investigate the influence of different ratios of style and content loss as suggested in~\citep{gatys2016image}. The results in Fig.~\ref{fig_csloss}~(b) indicate that an emphasis on content loss (lower style/content loss ratio) leads to better harmonization performance.
Despite the enhancements by DLEST+CS, it’s important to note that these come with a substantial rise in computational expense. This is primarily due to the computation of style and content loss in the ambient image space. Consequently, after each iteration of Stochastic Gradient Langevin Dynamics (SGLD), the latent code must be decoded. Furthermore, this process necessitates the use of an additional pre-trained feature extractor, adding to the complexity and computational demand.
%In real applications, it is important to consider the trade-off between computation efficiency and content preservation of MRIs. 

%the trade-off between using a faster latent content loss and employing a more robust separated style and content loss.

\if false  %暂时不加这个实验  May-14-2024
%%%%%%%%%%%%%%%%%%%% ComBat %%%%%%%%%%%%%%%%%%%%%%%%
%\subsection{Comparison with ComBat at Feature Level}
\subsection{Comparison with Feature-Level Harmonization}
Although the primary motivation behind proposing an image-level harmonization approach is its adaptability to a variety of downstream tasks, it remains crucial to compare our DLEST framework with the prevalent feature-level harmonization method, ComBat~\citep{ComBat}. 
This comparison generally follows the evaluation scheme outlined in the OpenBHB Challenge~\citep{OpenBHB} to ensure a fair assessment of both methodologies. 
Specifically, we employ the official Python implementation of ComBat \citep{neuroCombat,ComBat} to harmonize a collection of seven pre-extracted MRI features. 
These features include cortical thickness (along with standard deviation), GM volume, surface area, integrated mean (and Gaussian) curvature, and intrinsic curvature index for each subject in the training set. 
Subsequently, we train a linear logistic regression model~\citep{sklearn} on these harmonized features and evaluate its multi-class site classification performance on the validation set. 
Similarly, for the image-level approach, we utilize a VGG19 feature extractor (with weights frozen) to directly extract image features from the harmonized MRIs, and train a comparable linear logistic regression model for site classification. 
We also compare two baselines, named \textbf{Baseline-F} and \textbf{Baseline-I} that correspond to the original MRI features and images, respectively.
The accuracy and balanced accuracy, and F1 score are reported, with lower performance indicating more effective harmonization. 
As shown in Table~\ref{tab:linear_site_cl}, both the original MRI scans and extracted MRI features are characterized by a significant amount of site-related variations. 
ComBat performs better than Baseline-F, indicating its efficacy in feature harmonization. 
%But it is generally worse than learning-based methods. 
However, it only addresses the site effect on the pre-extracted features, rather than the entire image level, limiting its utility in broader applications.
%While ComBat is capable of effectively harmonizing the provided features, it still retains a higher number of site-related features compared to all image-level harmonization methods. 
In contrast, our method successfully eliminates the majority of the site-related variations, resulting in the lowest BACC for the linear model.

\begin{table}
    \centering
    \setlength{\tabcolsep}{4pt}
\caption{Site classification result of linear model where ‘*’ denotes the difference between a specific method and
DLEST is statistically significant (p $<$ 0.05 via paired t-test).}
\label{tab:linear_site_cl}
\scriptsize
    \begin{tabular}{l|ccc} 
        \toprule
         Method&  BACC$\downarrow$&  ACC$\downarrow$& F1$\downarrow$\\ 
         \midrule
         Baseline-F&  $0.703\pm 0.019^*$&  $0.762\pm 0.008^*$& $0.739\pm 0.020^*$\\ 
         Baseline-I   & $0.311\pm0.087^*$& $0.505\pm0.086^*$& $0.451\pm0.083^*$\\
         \textbf{ComBat}&  $0.586\pm 0.158^*$&  $0.566\pm 0.213^*$& $0.525\pm 0.230^*$\\ 
         CycleGAN&  $0.134\pm0.029$&  $0.392\pm0.060$& $0.303\pm0.046^*$\\ 
         StarGAN&  $0.178\pm0.053^*$&  $0.430\pm0.112$& $0.352\pm0.088^*$\\ 
         ImUnity&  $0.196\pm0.043^*$&  $0.406\pm0.107$& $0.350\pm0.080^*$\\ 
         \textbf{DLEST (Ours)}& $0.130\pm0.010$& $0.407\pm0.013$&$0.321\pm0.010$\\ 
        \bottomrule
    \end{tabular}

\end{table}
%%%%%%%%%%%%%%%%%%%%%%%%%%%%%%%%%%%%%%%%%%%%%%%%%%%%%%%%%%%%%%
\fi

\begin{figure}[t]
\setlength{\abovecaptionskip}{0pt}
\setlength{\belowcaptionskip}{-2pt}
\setlength{\abovedisplayskip}{-2pt}
\setlength{\belowdisplayskip}{-2pt}
\centering
\includegraphics[width=0.48\textwidth] {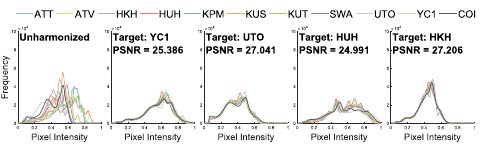}
\caption{Histogram of harmonized MRIs using four different target sites (\ie, YC1, UTO, HUH, and HKH) on SRPBS. Given a specific target site, MRIs from the remaining 10 sites are used as source data. The PSNR value denotes the intra-site variation of each target site.}
\label{fig_additional_sites}
\end{figure}

\subsection{Influence of Target Site}
\label{sec:target}
% Using intensity correlation variation 

% Minimal retraining - Emphasize this point! 
We further examine the impact of target site selection on harmonization using
% Our analysis compares intra-site variations across each site in the SRPBS dataset by computing the mean PSNR for all image pairs within a given site.
% Given that the SRPBS dataset encompasses all traveling subjects, each site shares an identical subject cohort, thereby containing consistent anatomical information. Consequently, a site yielding a higher intra-site PSNR score suggests a greater similarity in image style within that particular site.
four different target sites (\ie, YC1, UTO, HUH, and HKH) in the SRPBS dataset.
Similar to the main experiments, given a selected target site, we harmonize the remaining 10 sites as the source domain.
% {\color{blue}
% To investigate the influence of the target site selection, we perform another group of experiments by using four different target sites (\ie, XXX, ) from the SRPBS cohorts. 
% Similar to the main experiments, give a selected target site, we use the remaining 10 sites as the source domain.}  
Notably, only the SST module requires fine-tuning with each target change, typically requiring about 100 2D MRI slices from the target domain. 
This process incurs minimal computational cost, as detailed in Section~\ref{sec:cost}. The histogram results of the harmonized MRIs, depicted in Fig.~\ref{fig_additional_sites}, underscore the influence of target site selection on the harmonization outcome. When a target site with lower intra-site variations (higher PSNR), such as site HKH, is used, the harmonized histograms across all sites align more effectively. Conversely, a target site with a lower PSNR, such as site HUH, may result in less optimal histogram alignment. This observation suggests that a target site with excessive variations could hinder the model’s ability to accurately capture the latent data distribution that characterizes the target style.

\begin{table}[!t]
% \centering
% \setlength{\abovecaptionskip}{0pt}
% \setlength{\belowcaptionskip}{-2pt}
% \setlength{\abovedisplayskip}{-2pt}
% \setlength{\belowdisplayskip}{-2pt}
%\small
\caption{Results of time and computation cost. For DLEST, ``$a+b$'' denotes the number for SIG $+$ SST. The term ``Test time'' is the cost of generalizing on an unseen dataset. M: Million; GMac: Giga multiply-accumulate operations; H: Hour.}
%Site classification results produced by six different methods on the OpenBHB dataset (mean$\pm$standard deviation), where `*' denote the BACC difference between a specific competing method and the DLEST is statistically significant ($p\leq0.05$ via paired $t$-test).}%  with lower values indicating better results.} %removal of site-related variations.}
% \scriptsize
\footnotesize
% \tiny
\setlength{\tabcolsep}{2pt} % to reduce space between columns
% \begin{tabular}{l|ccccc}
\begin{tabular}{l|p{1.5cm}p{1cm}p{1cm}p{1cm}p{0.8cm}}
\toprule
%\multirow{2}{*}{Method} &\multicolumn{5}{c}{Metrics}\\ \cmidrule(l){2-6}
% \cmidrule(l){6-9}
% \cmidrule(l){10-13}
Method        & {$\#$Parameters (M)} & {FLOPs (GMac)}    & {Training Time (H)}  & {Test Time (H)}  & {Need Re-training}\\
\midrule
CycleGAN  & $28.3$ & $1794.1$   & $29.1$  & $8.0$  & $Yes$ \\
StyleGAN   & $161.3$ & $4865.3$  & $53.8$  & $9.1$  & $Yes$ \\
ImUnity  & $671.0$ & $246.2$ & $25.3$ & $1.5$ & $No$ \\
\textbf{DLEST~(Ours)}   & $21.7$+$5.3$ & $2.1$+$0.3$ & $20.0$+$0.4$ & $0.4$  & $No$ \\
\bottomrule
\end{tabular}
\label{tab:cost}
\end{table}

\subsection{Computational Cost Evaluation}
\label{sec:cost}
To quantitatively assess the enhancements in computational and time efficiency, we compare our DLEST framework with three GAN-based SOTA methods across several metrics. 
These include the number of trainable parameters, total floating point operations (FLOPs) per forward pass, training time to convergence on the OpenBHB training dataset, testing time on the SRPBS dataset, and the necessity for retraining when applied to an unseen dataset. 
As demonstrated in Table~\ref{tab:cost}, our model, encompassing both the SIG and SST modules, contains the fewest parameters, requires significantly fewer FLOPs per forward pass, and generally requires less training time. 
Furthermore, in comparison to ImUnity, which can also generalize to unseen data without retraining, our DLEST framework requires less test time to fine-tune solely the SST module. 
These results collectively suggest that our DLEST  can achieve highly efficient and generalizable MRI harmonization.

\if false
\begin{table}[!t]
% \centering
% \setlength{\abovecaptionskip}{0pt}
% \setlength{\belowcaptionskip}{-2pt}
% \setlength{\abovedisplayskip}{-2pt}
% \setlength{\belowdisplayskip}{-2pt}
%\small
\footnotesize
\caption{Time and computation cost. For  DLEST, $a+b$ denotes the number for SIG $+$ SST. Test time is the cost of generalizing on an unseen dataset. M: Million; GMac: Giga multiply–accumulate operations; H: Hour.}
%Site classification results produced by six different methods on the OpenBHB dataset (mean$\pm$standard deviation), where `*' denote the BACC difference between a specific competing method and the DLEST is statistically significant ($p\leq0.05$ via paired $t$-test).}%  with lower values indicating better results.} %removal of site-related variations.}
\scriptsize
%\footnotesize
\setlength{\tabcolsep}{2pt} % to reduce space between columns
% \begin{tabular}{l|ccccc}
\scriptsize
\begin{tabular}{l|p{1.7cm}p{1.5cm}p{1.2cm}p{1cm}p{1cm}}
\toprule
%\multirow{2}{*}{Method} &\multicolumn{5}{c}{Metrics}\\ \cmidrule(l){2-6}
% \cmidrule(l){6-9}
% \cmidrule(l){10-13}
Method        & {$\#$Parameters (M)} & {FLOPs (GMac)}    & {Training Time (H)}  & {Test Time (H)}  & {Need Re-training}\\
\midrule
CycleGAN  & $28.3$ & $1794.1$   & $29.1$  & $8.0$  & $Yes$ \\
StyleGAN   & $161.3$ & $4865.3$  & $53.8$  & $9.1$  & $Yes$ \\
ImUnity  & $671.0$ & $246.2$ & $25.3$ & $1.5$ & $No$ \\
\textbf{DLEST~(Ours)}   & $21.7$+$5.3$ & $2.1$+$0.3$ & $20.0$+$0.4$ & $0.4$  & $No$ \\
\bottomrule
\end{tabular}
\label{tab:cost}
\end{table}
\fi

\subsection{Limitations and Future Work}
While extensive experiments demonstrate that the proposed DLEST is competitive over SOTA methods, it is important to acknowledge some limitations. % and identify directions for future research.  
\emph{First}, our current implementation focuses on 2D-level MRI slices, potentially overlooking volumetric variations, and can only synthesize separate MRI slices instead of the whole volume. 
%To address this, one possible solution is to extend our implementation to 2.5D, incorporating a conditioning mechanism in the autoencoder to generate MRI slices by conditioning on adjacent slices~\citep{3D_diffusiton_brain}. 
One possible solution %to address is to limitation is 
is to extend our implementation to 2.5D, incorporating a conditioning mechanism in the autoencoder to generate MRI slices by conditioning on adjacent slices~\citep{peng2022generating}.
% As our framework disentangles the image generation step and the style translation step, such extension can be achieved by only adding a conditioning mechanism to the autoencoder in the first part while the energy-based style translation model remains unchanged.} 
\emph{Second}, our current autoencoder employs a one-dimensional latent code to achieve the most efficient encoding and style translation. 
However, the aggressive 1D encoding discards some inherent spatial structures of the original image, leading to some minor loss of anatomical structures during reconstruction. 
%{\color{red}
%Recent research~\citep{rombach2022high}} suggests 
It is interesting to employ %using a 
2D latent encoding~\citep{rombach2022high} with a relatively mild compression rate to better preserve anatomical structures of input MRIs, which will be one of our future works. 
\emph{In addition}, our current work only uses latent information from images, without considering biological features, such as gender, age, and relevant clinical diagnostic scores. 
As another further work, we plan to incorporate biological variables as a condition to improve the fidelity of generated images. %when performing image-level harmonization.}
%{\color{blue}\emph{Furthermore}, Can we add some discussion on another future direction by explicitly incorporating biological variables (\ie, covariates) into latent code space?}

%{\color{red}
%Please use ``citep'' other than ``citep''}

\section{Conclusion}
This paper proposes a novel unpaired image-level harmonization framework for multi-site MRI data. 
By disentangling site-invariant image generation and site-specific style translation, our model can generalize on new sites without any retraining on original data. 
The style translation in latent space efficiently harmonizes multiple source sites to the target site. %, even with minimal target samples. 
Moreover, the model can synthesize diverse MRIs in a site-specific style without additional training. 
Quantitative and qualitative evaluations suggest that our DLEST outperforms several state-of-the-art harmonization methods. 
Additionally, our framework can be employed as a pre-processing step in downstream studies, such as mitigating site effects in multi-site MRIs and synthesizing MRIs to train general models.

\vspace{-4pt}
\bibliographystyle{model2-names.bst}
\biboptions{authoryear}
\bibliography{TMI2022}

\end{document}

%% file: defs.tex
\def\x{{\bf x}}
\def\y{{\bf y}}

\def\0{{\bf 0}}
\def\1{{\bf 1}}

\def\etal{{\em et al.}}
\def\eg{{\em e.g.}}
\def\ie{{\em i.e.}}

\def\etal{{\em et al.\/}\,}